\documentclass[9pt]{IEEEtran}
\usepackage{amsmath,amsfonts}
\usepackage{amsmath}
\usepackage{algorithmic}
\usepackage{array}
\usepackage[caption=false,font=normalsize,labelfont=sf,textfont=sf]{subfig}
\usepackage{textcomp}
\usepackage{stfloats}
\usepackage{url}
\usepackage{verbatim}
\usepackage{graphicx}
\usepackage{algorithm}
\usepackage{setspace}
\usepackage{color}
\usepackage{cite}
\usepackage{booktabs}
\usepackage{float}
\newtheorem{remark}{\bf \ \ Remark}

\newtheorem{theorem}{\bf \ \ Theorem}
\newtheorem{lemma}{\bf \ \ Lemma}
\newtheorem{assumption}{\bf \ \ Assumption}
\usepackage{textcomp}
\def\BibTeX{{\rm B\kern-.05em{\sc i\kern-.025em b}\kern-.08em
    T\kern-.1667em\lower.7ex\hbox{E}\kern-.125emX}}
\begin{document}
\title{Cooperative Optimal Output Tracking for Discrete-Time Multiagent Systems: Stabilizing Policy Iteration Frameworks}
\author{Dongdong Li, Jiuxiang Dong
\thanks{This work was supported in part by the National Natural Science Foundation of China under Grant 62273079 and in part by the Research Fund of State Key Laboratory of Synthetical Automation for Process Industries in China under Grant 2013ZCX01.
 ({\it Corresponding author: Jiuxiang Dong})}
\thanks{Dongdong Li and Jiuxiang Dong are with the College of Information Science and Engineering, and
the State Key Laboratory of Synthetical Automation of Process Industries, Northeastern University, Shenyang 110819, China. Email: lidongdongyq@163.com, dongjiuxiang@ise.neu.edu.cn.}}

\maketitle

\begin{abstract}
This paper proposes two cooperative optimal output tracking (COOT) algorithms based on policy iteration (PI) for discrete-time multi-agent systems with unknown model parameters. First, we establish a stabilizing PI framework that can start from any initial control policy, relaxing the dependence of traditional PI on the initial stabilizing control policy. Then, another efficient and equivalent $Q$-learning framework is developed, which is shown to require only less system data to get the same results as the stabilizing PI. In the two frameworks, the stabilizing control policy is obtained by gradually iterating the stabilizing virtual system to the actual feedback closed-loop system. Two explicit schemes for adjusting the iteration step-size/coefficient are designed and their stability is analyzed. Finally, the COOT is realized by a distributed feedforward-feedback controller with learned optimal gains. The proposed algorithms are validated by simulation.
\end{abstract}


\section{Introduction}
Reinforcement learning (RL) \cite{lewis2010reinforcement} or adaptive dynamic programming \cite{lewis2012optimal}, which makes decisions through the interaction of the agent with the environment, can combine features of optimal control and intelligent computation.
The $Q$-learning algorithm is one of the classical techniques in RL and is usually designed by means of an action-valued function, which is defined by the Bellman equation and is often referred to as the $Q$-function \cite{luo2020policy,li2024data}. Most RL results, including $Q$-learning, for discrete-time (DT) systems or continuous-time (CT) systems are designed based on two main frameworks: policy iteration (PI) \cite{hewer1971iterative,lewis2010reinforcement,chen2022robust,kiumarsi2014reinforcement,kiumarsi2015optimal,jiang2019optimal,luo2020policy,lopez2023efficient,al2007model} and value iteration (VI) \cite{lewis2010reinforcement,kiumarsi2015optimal,li2022model}.
Recently, RL algorithms have been applied to the cooperative optimal output tracking (COOT) problem for multi-agent systems (MASs) and many classical results have been obtained such as \cite{gao2018leader,chen2020off,jiang2023reinforcement,chen2023distributed,xie2023data}. Among these methods, COOT is mainly realized via the combination of output regulation theory \cite{huang2004nonlinear} and RL. Considering the communication topology, the output regulation/tracking is mainly solved by internal model principle \cite{gao2018leader,chen2020off,chen2023distributed,xie2023data} or solving the regulator equations \cite{jiang2023reinforcement}, while the optimal control is mainly realized by PI \cite{gao2018leader,chen2020off,chen2023distributed} and VI \cite{jiang2023reinforcement,xie2023data}.

It is worth noting that the PI starts from a stabilizing control policy and gradually improves the control policy by solving a series of Lyapunnov functions, which is characterized by fast convergence \cite{kiumarsi2015optimal,luo2020policy,li2024data}. VI can start with an arbitrary control policy but the iterations converge more slowly \cite{luo2020policy,li2024data}.
Recently, many effective methods have been investigated to address the initial stabilizing control limitations of PI for DT \cite{luo2019balancing,lamperski2020computing,li2024data} and CT \cite{chen2022homotopic,jiang2022bias,gao2022resilient} systems. In \cite{luo2019balancing,gao2022resilient}, the advantages of VI are utilized to address the initial stabilizing control limitations of PI. In \cite{luo2019balancing}, the coefficients are used to balance VI and PI, while in \cite{gao2022resilient,10167108}, the stabilizing control gain is computed from VI and then used for PI to get the optimal control quickly, which is called hybrid iteration. In \cite{jiang2022bias}, a bias-PI algorithm is proposed for CT systems that does not require an initial stabilizing policy. Recently, some methods for calculating directly stabilizing control policy have been proposed for DT \cite{lamperski2020computing,li2024data,lopez2023efficient} and CT \cite{chen2022homotopic} systems. In \cite{lopez2023efficient}, a deadbeat control gain matrix is designed to initialize the proposed efficient off-policy $Q$-learning algorithm. In \cite{lamperski2020computing}, a discount factor-based PI method is designed to achieve optimal control without an initial stabilizing policy. In \cite{li2024data}, a stabilizing optimal control algorithm starting from a zero-initial control policy is presented. In \cite{chen2022homotopic,li2024off}, homotopy-based PI algorithms are proposed to obtain the stabilizing control policy for CT linear systems by designing the stabilizing system and introducing the homotopy parameters. These methods are difficult to directly extend to DT systems, but inspired the work in this paper.

Inspired by above results, we propose two distributed COOT algorithms. The main contributions are as follows:
(i). A RL-based distributed COOT framework with the feedforward-feedback controller is proposed, where the feedforward and feedback policies are learned by two different online iteration methods; (ii). Compared with \cite{li2024data,hewer1971iterative,chen2022robust,kiumarsi2014reinforcement,kiumarsi2015optimal,jiang2019optimal,al2007model}, we develop a novel Lyapunov-based stabilizing PI (SPI) that can start with an arbitrary initial control policy, and obtain the stabilizing control policy by building the stabilizing virtual closed-loop system and iterating it to the actual closed-loop system. An explicit scheme for adjusting the iteration step-size is designed to ensure the stability of the virtual system at each iteration step; (iii). Compared with $Q$-learning methods \cite{kiumarsi2014reinforcement,luo2020policy,lopez2023efficient}, we extend the SPI to $Q$-learning, which provides a new perspective for computing the stabilizing control policy and achieving optimal control. Another explicit iteration step-size selection scheme is designed based on the stability equation for the proposed $Q$-learning, which ensures the stability of the virtual system. Moreover, a low data full column-rank condition is designed by separating the policy evaluation of $Q$-learning from the process of solving the output regulator equations.

{\it Notations:}
For any matrix $Y\in\mathbb{R}^{a\times b}$, define $\mathrm{vec}(Y)=[Y_{1}^{\top},Y_{2}^{\top},\ldots,Y_{b}^{\top}]^{\top}\in\mathbb{R}^{ab}$, where $Y_{j}\in\mathbb{R}^{a}$ is $i$th column of matrix $Y$ for $j=1,\ldots,b$. If $Y=Y^{\top}\in\mathbb{R}^{a\times a}$, $\mathrm{vecs}(Y)=[Y_{1,1},2Y_{1,2},\ldots,2Y_{1,a},Y_{2,2},2Y_{2,3},\ldots,2Y_{a-1,a},Y_{a,a}]^{\top}\in\mathbb{R}^{\frac{a(a+1)}{2}}$, where $Y_{i,j}$ is the $i$th row and $j$th column element of matrix $Y$. For vector $v\in\mathbb{R}^{a}$, $\mathrm{vecv}(v)=[v_{1}^{2},v_{1}v_{2},\ldots,
v_{1}v_{n},v_{2}^{2},v_{2}v_{3},\ldots,v_{a-1}v_{a},v_{a}^{2}]^{\top}\in\mathbb{R}^{\frac{a(a+1)}{2}}$. $\sigma_{\max}(\ast)$ and $\sigma_{\min}(\ast)$ denote the maximum and minimum singular values, and $\rho(\ast)$ denotes the spectral radius.  A directed graph is defined as $\bar{\mathcal{G}}=(\bar{\mathcal{V}},\bar{\mathcal{E}})$ with $\bar{\mathcal{E}}\subseteq\bar{\mathcal{V}}\times\bar{\mathcal{V}}$ and $\bar{\mathcal{V}}=\{0,1,2,\ldots,{N}\}$. Node $0$ denotes the leader, $i=1,2,\ldots,{N}$ nodes denote followers.
Define $\mathcal{H}=[h_{ij}]_{i,j=1}^{N}\in\mathbb{R}^{{N}\times{N}}$, where $h_{ii}=\sum_{j=0}^{N}a_{ij}$ and $h_{ij}=-a_{ij}$ with $a_{ij}>0$ if $(j,i)\in\bar{\mathcal{E}}$, otherwise $a_{ij}=0$.

\section{Problem Formulation And Preliminaries}\label{section:2}
A class of linear heterogeneous DT MASs is described as follows
\begin{align}
{x}_{i}(t+1)&=A_{i}x_{i}(t)+B_{i}u_{i}(t),\quad  i=1,2,\ldots,{N},\label{1a}\\
y_{i}(t)&=C_{i}x_{i}(t)+S_{i}u_{i}(t), \label{1b}\\
e_{i}(t)&=C_{i}x_{i}(t)+S_{i}u_{i}(t)+Fv(t),\label{1c}
\end{align}
where $x_{i}(t)\in\mathbb{R}^{n_i}$, $u_{i}(t)\in\mathbb{R}^{m_i}$ and $y_{i}(t)\in\mathbb{R}^{n_{y}}$ are the state, input and output of agent $i$, and $e_{i}(t)\in\mathbb{R}^{n_{y}}$ is the output tracking error of agent $i$ for the leader. $v(t)\in\mathbb{R}^{n_{v}}$ is the leader state as
\begin{equation}\label{2}
\begin{aligned}
v(t+1)=Ev(t),\quad y_{d}(t)=-Fv(t),
\end{aligned}
\end{equation}
where $y_{d}\in\mathbb{R}^{y}$ is its output.
 $A_{i}\in\mathbb{R}^{n_i\times n_i}$, $B_{i}\in\mathbb{R}^{n_i\times m_i}$, $C_{i}\in\mathbb{R}^{n_{y}\times n_i}$, $S_{i}\in\mathbb{R}^{n_{y}\times m_i}$, $E\in\mathbb{R}^{n_{v}\times n_{v}}$ and $F\in\mathbb{R}^{n_{y}\times n_v}$ are constant matrices, where $A_{i}$ and $B_{i}$ are unknown.

\begin{assumption}\cite{huang2016cooperative}\label{A1}
The $(A_{i}, B_{i})$ is stabilizable.
\end{assumption}

\begin{assumption}\cite{huang2016cooperative}\label{A2}
$\mathrm{rank}(\left[
\begin{array}{cc}
   A_{i}-\lambda I& B_{i} \\
   C_{i} & S_{i}
\end{array}
\right])=n_{i}+n_{y}$ for $i=1,\ldots,N$, where $\forall\lambda\in\sigma(E)$.
\end{assumption}

\begin{assumption}\cite{huang2016cooperative}\label{A3}
Graph $\bar{\mathcal{G}}$ contains a directed spanning tree.
\end{assumption}

\begin{assumption}\cite{huang2016cooperative}\label{A4}
All eigenvalue modulus of matrix $E$ are less than or equal to 1.
\end{assumption}

The COOT is to design the distributed controllers for all agents such that the output tracking error $e_{i}(t)$ converges asymptotically to 0 and the predefined performances are optimized.
We design the following observer to estimate $E$, $F$ and $v(t)$ as
\begin{subequations}\label{3}
\begin{align}
&E_{i}(t+1)=E_{i}(t)+\mu_{i}\sum_{j=0}^{N}a_{ij}(E_{j}(t)-E_{i}(t)),\label{3a}\\
&F_{i}(t+1)=F_{i}(t)+\mu_{i}\sum_{j=0}^{N}a_{ij}(F_{j}(t)-F_{i}(t)),\label{3b}\\
&\zeta_{i}(t+1)=E_{i}(t)\zeta_{i}(t)+\mu_{i}E_{i}(t)\sum_{j=0}^{N}a_{ij}(\zeta_{j}(t)-\zeta_{i}(t)),\label{3c}
\end{align}
\end{subequations}
where $E_{i}\in\mathbb{R}^{n_{v}\times n_{v}}$, $F_{i}\in\mathbb{R}^{n_{y}\times n_{v}}$ and $\zeta_{i}\in\mathbb{R}^{n_{v}}$ are the estimated values of $E$, $F$ and $v(t)$ with $E_{0}=E$, $F_{0}=F$ and $\zeta_{0}=v$. Define the estimated errors as $\tilde{E}_{i}={E}_{i}-E$, $\tilde{F}_{i}={F}_{i}-F$ and $\tilde{\zeta}_{i}={\zeta}_{i}-v$. 
\begin{lemma}\label{L1}
under Assumption \ref{A4}, given the observer \eqref{3}. For any initial $E_{i}(0)$, $F_{i}(0)$ and $\zeta_{i}(0)$ with $i=1,2,\ldots,N$, if $0<\mu_{i}<2/\rho(\mathcal{H})$, one has that $\lim_{t\rightarrow\infty}\tilde{E}_{i}(t)=0$, $\lim_{t\rightarrow\infty}\tilde{F}_{i}(t)=0$ and $\lim_{t\rightarrow\infty}\tilde{\zeta}_{i}(t)=0$ exponentially.
\end{lemma}

{\it Proof.}
Under Assumption \ref{A4}, the proof that $\lim_{t\rightarrow\infty}\tilde{E}_{i}(t)=0$ and $\lim_{t\rightarrow\infty}\tilde{\zeta}_{i}(t)=0$ exponentially is the same as \cite[Lem. 2]{huang2016cooperative} with $\rho(E)\leq1$. From \eqref{3b}, one gets $\tilde{F}(t+1)=(\mathcal{H}^{\mu}\otimes I_{n_{y}})\tilde{F}(t)$, where $\tilde{F}:=\mathrm{col}(\tilde{F}_{1},\ldots,\tilde{F}_{N})$ and $\mathcal{H}^{\mu}:=I_{N}-\mathrm{diag}\{\mu_{i}\}\mathcal{H}$. Thus, if $0<\mu_{i}<2/\rho(\mathcal{H})$, then $\lim_{t\rightarrow\infty}\tilde{F}_{i}(t)=0$ exponentially.
$\Box$

The feedforward-feedback controller is designed as
\begin{equation}\label{4}
\begin{aligned}
u_{i}(t)=-K_{i}x_{i}(t)+T_{i}\zeta_{i}(t),
\end{aligned}
\end{equation}
where $K_{i}\in\mathbb{R}^{m_{i}\times n_{i}}$ is the feedback control gain and $T_{i}\in\mathbb{R}^{m_{i}\times n_{v}}$ is the feedforward control gain. According to output regulation theory \cite{huang2004nonlinear,huang2016cooperative}, $T_{i}$ can be obtained by
\begin{equation}\label{5}
\begin{aligned}
T_{i}=U_{i}+K_{i}X_{i},
\end{aligned}
\end{equation}
where pair $(X_{i}\in\mathbb{R}^{n_{i}\times n_{v}}, U_{i}\in\mathbb{R}^{m_{i}\times n_{v}})$ is the solution to the following regulator equations,
\begin{subequations}\label{6}
\begin{align}
0&=A_{i}X_{i}+B_{i}U_{i}-X_{i}E,\label{6a}\\
0&=C_{i}X_{i}+S_{i}U_{i}+F.\label{6b}
\end{align}
\end{subequations}

\begin{lemma}\label{L2}
Consider system $x(t+1)=f_{1}x(t)+f_{2}(t)$, where $x\in\mathbb{R}^{n}$, $f_{1}\in\mathbb{R}^{n\times n}$ is Schur, and $f_{2}\in\mathbb{R}^{n}$ is bounded for $t>0$. Then, for any $x(0)$, there is $\lim_{t\rightarrow\infty}x(t)=0$ if $\lim_{t\rightarrow\infty}f_{2}(t)=0$.
\end{lemma}

{\it Proof.}
The proof is similar to \cite[Lem. 1]{huang2016cooperative}, is omitted here. $\Box$

\begin{theorem}\label{the1}
Under Assumptions \ref{A1}-\ref{A4}, given MASs \eqref{1a}-\eqref{2}. If $T_{i}$ is designed by \eqref{5}-\eqref{6b} and $K_{i}$ is designed such that $A_{i}-B_{i}K_{i}$ is Schur for $i=1,\ldots,N$, then the cooperative output tracking can be realized by the controller \eqref{4}.
\end{theorem}

{\it Proof.}
Define $\bar{x}_{i}(t)=x_{i}(t)-X_{i}v(t)$. By using \eqref{1a}-\eqref{2}, \eqref{4}-\eqref{6} and $\tilde{\zeta}_{i}(t)=\zeta_{i}(t)-v(t)$, we have
$\bar{x}_{i}(t+1)=A_{i}x_{i}(t)+B_{i}u_{i}(t)-X_{i}Ev(t)=(A_{i}-B_{i}K_{i})\bar{x}_{i}(t)+B_{i}T_{i}\tilde{\zeta}_{i}(t)$,
where $\lim_{t\rightarrow\infty}\tilde{\zeta}_{i}(t)=0$ exponentially by Lemma \ref{L1}. By Lemma \ref{L2}, there is $\lim_{t\rightarrow\infty}\bar{x}_{i}(t)=0$ due to $A_{i}-B_{i}K_{i}$ is Schur. Since $e_{i}(t)=(C_{i}-S_{i}K_{i})\bar{x}_{i}(t)+S_{i}T_{i}\tilde{\zeta}_{i}(t)$, one has $\lim_{t\rightarrow\infty}e_{i}(t)=0$.
$\Box$

By Defining $\bar{u}_{i}(t)=u_{i}-U_{i}v(t)$, we obtain the error system as
\begin{subequations}\label{8}
\begin{align}
\bar{x}_{i}(t+1)&=A_{i}\bar{x}_{i}(t)+B_{i}\bar{u}_{i}(t),\label{8a}\\
e_{i}(t)&=C_{i}\bar{x}_{i}(t)+S_{i}\bar{u}_{i}(t).\label{8b}
\end{align}
\end{subequations}
For $i=1,\ldots,N$, the optimal feedback controller is designed as
\begin{equation}\label{9}
\begin{aligned}
\bar{u}_{i}(t)=-K_{i}^*\bar{x}_{i}(t),
\end{aligned}
\end{equation}
where $K_{i}^*$ is the optimal feedback gain such that the following performance is minimized,
\begin{equation}\label{10}
\begin{aligned}
\min_{\bar{u}_{i}}&\{V_{i}(t)=\sum_{\tau=t}^{\infty}(\bar{x}_{i}^{\top}(\tau)Q_{i}\bar{x}_{i}(\tau)+\bar{u}_{i}^{\top}(\tau)R_{i}\bar{u}_{i}(\tau))\},
\end{aligned}
\end{equation}
with $Q_{i}=Q_{i}^{\top}>0$ and $R_{i}=R_{i}^{\top}>0$. If $K_{i}^*$ is obtained by solving problem \eqref{10} and $(X_{i}, U_{i})$ is obtained by solving the output regulator equations \eqref{6a}-\eqref{6b}, respectively, then the COOT problem is solved.

The optimization problem in \eqref{10} can be solved by designing the optimal feedback gain as
\begin{equation}\label{11}
\begin{aligned}
K_{i}^*=(R_{i}+B_{i}^{\top}P_{i}B_{i})^{-1}B_{i}^{\top}P_{i}A_{i}
\end{aligned}
\end{equation}
where $P_{i}=P_{i}^{\top}\in\mathbb{R}^{n_{i}\times n_{i}}$ is the unique positive definite solution to the
Algebraic Riccati Equation (ARE),
\begin{equation}\label{12}
\begin{aligned}
P_{i}=A_{i}^{\top}P_{i}A_{i}-A_{i}^{\top}P_{i}B_{i}(R_{i}+B_{i}^{\top}P_{i}B_{i})^{-1}B_{i}^{\top}P_{i}A_{i}+Q_{i}.
\end{aligned}
\end{equation}
\begin{lemma}\label{L3}\cite{hewer1971iterative}
Given initial stabilizing control gain $K_{i}^{0}$ such that $\rho(A_{i}-B_{i}K_{i}^{0})<1$. Find $P_{i}^{j}=(P_{i}^{j})^{\top}$ by Lyapunov equation,
\begin{equation}\label{13}
\begin{aligned}
P_{i}^{j}=(A_{i}-B_{i}K_{i}^{j})^{\top}P_{i}^{j}(A_{i}-B_{i}K_{i}^{j})+Q_{i}+(K_{i}^{j})^{\top}R_{i}K_{i}^{j}
\end{aligned}
\end{equation}
and update the policy by
\begin{equation}\label{14}
\begin{aligned}
K_{i}^{j+1}=(R_{i}+B_{i}^{\top}P_{i}B_{i})^{-1}B_{i}^{\top}P_{i}^{j}A_{i}
\end{aligned}
\end{equation}
for $j=0,1,2,\ldots$ and $i=1,\ldots,N$. Then, there are: 1). $\rho(A_{i}-B_{i}K_{i}^{j+1})<1$; 2). $P_{i}^*\leq P_{i}^{j+1}\leq P_{i}^{j}$; 3). $\lim_{j\rightarrow\infty}P_{i}^{j}=P_{i}^*$ and $\lim_{j\rightarrow\infty}K_{i}^{j}=K_{i}^*$.
\end{lemma}

Lemma \ref{L3} is the classical Lyapunov-based PI algorithm, which can be expanded as data-driven versions \cite{chen2022robust,jiang2019optimal,kiumarsi2015optimal} to solve ARE \eqref{12}. However, Lemma \ref{L3} requires a known initial stabilizing policy.

\section{Model-Based Solution to COOT Problem} \label{section:3}

\subsection{Model-Based Solution to Regulator Equations} \label{section:3:1}
Inspired by \cite{jiang2023reinforcement,jiang2024adaptive}, we define two Sylvester maps $\Upsilon_{i}:\mathbb{R}^{n_{i}\times n_{v}}\rightarrow\mathbb{R}^{n_{i}\times n_{v}}$ and $\bar{\Upsilon}_{i}: \mathbb{R}^{n_{i}\times n_{v}}\times\mathbb{R}^{m_{i}\times n_{v}}\rightarrow\mathbb{R}^{n_{i}\times n_{v}}$ as
\begin{subequations}\label{15}
\begin{align}
\Upsilon_{i}(X_{i}):=&X_{i}E-A_{i}X_{i},\label{15a}\\
\bar{\Upsilon}_{i}(X_{i},U_{i}):=&X_{i}E-A_{i}X_{i}-B_{i}U_{i}.\label{15b}
\end{align}
\end{subequations}
Then, the solution to \eqref{6a}-\eqref{6b} can be established by the Sylvester maps. Set sequences $X_{il}\in\mathbb{R}^{n_{i}\times n_{v}}$ and $U_{il}\in\mathbb{R}^{m_{i}\times n_{v}}$ for $l=0,1,\ldots,h_{i}$ and $i=1,\ldots,N$, where $(h_{i}-1)$ is the dimension of the null space
of $I_{n_{v}}\otimes \bar{C}_{i}$ with $\bar{C}_{i}=[C_{i},S_{i}]$ and $\otimes$ being the Kronecker product. We select $X_{i0}=0_{n_{i}\times n_v}$, $U_{i0}=0_{m_{i}\times n_v}$, all the vectors $\mathrm{vec}([X_{il}^{\top},U_{il}^{\top}])$ to form a basis for $\mathrm{ker}(I_{n_{v}}\otimes\bar{C}_{i})$ for $l=2,\ldots,h_{i}$, and $C_{i}X_{i1}+S_{i}U_{i1}=-F$. Note that $F$ is not available for all agents. Therefore, we design
\begin{equation}\label{16}
\begin{aligned}
&[\hat{X}_{i1}^{\top},\hat{U}_{i1}^{\top}]^{\top}=-\bar{C}_{i}^{\top}(\bar{C}_{i}\bar{C}_{i}^{\top})^{-1}F_{i}(t_{0}),
\end{aligned}
\end{equation}
where $t_{0}>0$ is a constant, $\hat{X}_{i1}$ and $\hat{U}_{i1}$ are the estimations of $X_{i1}$ and $U_{i1}$ with $\tilde{X}_{i1}=\hat{X}_{i1}-X_{i1}$ and $\tilde{U}_{i1}=\hat{U}_{i1}-U_{i1}$ being the estimation errors. Assumption \ref{A2} guarantees that $\bar{C}_{i}$ is full row-rank and hence $\bar{C}_{i}\bar{C}_{i}^{\top}$ is invertible. Since $[\tilde{X}_{i1}^{\top},\tilde{U}_{i1}^{\top}]^{\top}=-\bar{C}_{i}^{\top}(\bar{C}_{i}\bar{C}_{i}^{\top})^{-1}\tilde{F}_{i}(t_{0})$ and $\tilde{F}_{i}(t_{0})$ can be set sufficiently small from Lemma \ref{L1}, there is $(\tilde{X}_{i1},\tilde{U}_{i1})\rightarrow0$.

The solution to \eqref{6a}-\eqref{6b} can be approximated as
$(\hat{X}_{i},\hat{U}_{i})=(\hat{X}_{i1},\hat{U}_{i1})+\sum_{l=2}^{h_{i}}\delta_{il}(X_{il},U_{il})$,
where $\delta_{il}$ is the unknown coefficients. Unlike the CT systems \cite{jiang2023reinforcement}, by introducing arbitrary constant matrix $M_{i}\in\mathbb{R}^{n_{i}\times n_{i}}$, we have
$M_{i}\bar{\Upsilon}_{i}(\hat{X}_{i},\hat{U}_{i})=M_{i}\bar{\Upsilon}_{i}(\hat{X}_{i1},\hat{U}_{i1})+\sum_{l=2}^{h_{i}}\delta_{il}M_{i}\bar{\Upsilon}_{i}(X_{il},U_{il}).$
Then, the regulator equations \eqref{6} and its approximate version can be expressed as
\begin{subequations}\label{19}
\begin{align}
&\Omega_{i}\chi_{i}={\eta}_{i}, \label{19a}\\
&\Omega_{i}\hat{\chi}_{i}=\hat{\eta}_{i},\quad i=1,\ldots,N,\label{19b}
\end{align}
\end{subequations}
where
\begin{equation}
\begin{aligned}\nonumber
&\Omega_{i}=\\
&\left[
\begin{array}{cccc}
\mathrm{vec}(M_{i}\bar{\Upsilon}_{i}({X}_{i2},{U}_{i2}))&\ldots&\mathrm{vec}(M_{i}\bar{\Upsilon}_{i}({X}_{ih_{i}},{U}_{ih_{i}}))&0 \\
\mathrm{vec}([{X}_{i2}^{\top},{U}_{i2}^{\top}]^{\top})&\ldots&\mathrm{vec}([{X}_{ih_{i}}^{\top},{U}_{ih_{i}}^{\top}]^{\top})&-I
\end{array}
\right],\\
&\chi_{i}=\left[
\begin{array}{cccc}
\delta_{i2}&\ldots&\delta_{ih_{i}}&\mathrm{vec}([{X}_{i}^{\top},{U}_{i}^{\top})]^{\top}
\end{array}
\right]^{\top}\\
&\hat{\chi}_{i}=\left[
\begin{array}{cccc}
\delta_{i2}&\ldots&\delta_{ih_{i}}&\mathrm{vec}([\hat{X}_{i}^{\top},\hat{U}_{i}^{\top})]^{\top}
\end{array}
\right]^{\top},\quad \hat{\eta}_{i}=\\
&\left[
\begin{array}{c}
-\mathrm{vec}(M_{i}\bar{\Upsilon}_{i}(\hat{X}_{i1},\hat{U}_{i1}))\\
-\mathrm{vec}([\hat{X}_{i1}^{\top},\hat{U}_{i1}^{\top}]^{\top})
\end{array}
\right],
\eta_{i}=\left[
\begin{array}{c}
-\mathrm{vec}(M_{i}\bar{\Upsilon}_{i}({X}_{i1},{U}_{i1}))\\
-\mathrm{vec}([{X}_{i1}^{\top},{U}_{i1}^{\top}]^{\top})
\end{array}
\right]
\end{aligned}
\end{equation}
with $\tilde{\eta}_{i}=\hat{\eta}_{i}-\eta_{i}$ being the approximation error for $\eta_{i}$.

Inspired by \cite{huang2016cooperative,jiang2024adaptive}, we design the following iteration method to update the approximate solution to regulator equations \eqref{6a}-\eqref{6b} as
\begin{equation}\label{20}
\begin{aligned}
\hat{\chi}_{i}^{n+1}=\hat{\chi}_{i}^{n}-\kappa_{i}\Omega_{i}^{\top}(\Omega_{i}\hat{\chi}_{i}^{n}-\hat{\eta}_{i}),
\end{aligned}
\end{equation}
where $n=0,1,2,\ldots$. 

\begin{theorem} \label{the2}
Consider the distributed observer \eqref{6} and iteration equation \eqref{20}. Under Assumption 2, $i=1,\ldots,N$ and $n=0,1,2,\ldots\rightarrow\infty$, setting $0<\kappa_{i}<2/\rho(\Omega_{i}^{\top}\Omega_{i})$, the solution $\hat{\chi}_{i}^{n}$ from \eqref{20} converges to the solution $\chi_{i}^*$ of \eqref{19a}.
\end{theorem}

{\it Proof.}
 Under Assumption \ref{A2}, there is a vector $\chi_{i}^*$ satisfying $\Omega_{i}\chi_{i}^*=\eta_{i}$, where $\Omega_{i}\in\mathbb{R}^{n_{1}^{i}\times n_{2}^{i}}$ with $n_{1}^{i}:=(2n_{i}+m_{i})n_{v}$ and $n_{2}^{i}:=(2n_{i}+2m_{i}-n_{y})n_{v}$. It can be inferred from Assumption 2 that $m_{i}\geq n_{y}$, thus $n_{2}^{i}\geq n_{1}^{i}$. From Assumption 2, we have $0<\mathrm{rank}(\Omega_{i}):=n_{3}^{i}\leq n_{1}^{i}$. By singular value decomposition, there is an orthogonal matrix $W_{i}\in\mathbb{R}^{n_{2}^{i}\times n_{2}^{i}}$ and a full-column rank matrix $\bar{\Omega}_{i}\in\mathbb{R}^{n_{1}^{i}\times n_{3}^{i}}$ with $W_{i}^{\top}W_{i}=W_{i}W_{i}^{\top}=I$ so that $\Omega_{i}W_{i}=[\bar{\Omega}_{i},0]$,
\begin{equation}
\begin{aligned}\nonumber
&W_{i}^{\top}\Omega_{i}^{\top}\Omega_{i}W_{i}=\left[
\begin{array}{cc}
\bar{\Omega}_{i}^{\top}\bar{\Omega}_{i}&0\\
0&0
\end{array}
\right], W_{i}^{\top}\Omega_{i}^{\top}\eta_{i}=\left[
\begin{array}{c}
\bar{\Omega}_{i}^{\top}\eta_{i}\\
0
\end{array}
\right],
\end{aligned}
\end{equation}
and $W_{i}^{\top}\Omega_{i}^{\top}\hat{\eta}_{i}=[(\bar{\Omega}_{i}^{\top}\hat{\eta}_{i})^{\top}, 0^{\top}]^{\top}$. 
Define $\bar{\chi}_{i}^*=W_{i}^{\top}\chi_{i}^*=[\bar{\chi}_{i1}^{*\top},\bar{\chi}_{i2}^{*\top}]^{\top}$ and $\hat{\bar{\chi}}_{i}=[\hat{\bar{\chi}}_{i1}^{\top},\hat{\bar{\chi}}_{i2}^{\top}]^{\top}$, where $\bar{\chi}_{i1}^*\in\mathbb{R}^{n_{3}^{i}}$ and $\bar{\chi}_{i2}^*\in\mathbb{R}^{n_{2}^i-n_{3}^{i}}$. Then, $\bar{\Omega}_{i}\bar{\chi}^*_{i1}=\eta_{i}$ has unique solution.
Defining $\tilde{\chi}^{n}_{i1}=\hat{\bar{\chi}}^{n}_{i1}-\bar\chi_{i1}^*$ and $\tilde{\chi}^{n}_{i2}=\hat{\bar{\chi}}^{n}_{i2}-\bar{\chi}_{i2}^*$, we have
\begin{subequations}\label{21}
\begin{align}
&\tilde{\chi}^{n+1}_{i1}=(I-\kappa\bar{\Omega}_{i}^{\top}\bar{\Omega}_{i})\tilde{\chi}^{n}_{i1}+\kappa\bar{\Omega}_{i}^{\top}\tilde{\eta}_{i}, \label{21a}\\
&\tilde{\chi}^{n+1}_{i2}=\tilde{\chi}^{n}_{i2},\label{21b}
\end{align}
\end{subequations}
where $\tilde{\eta}_{i}=[-\mathrm{vec}(M_{i}\bar{\Upsilon}_{i}(\tilde{X}_{i1},\tilde{U}_{i1}))^{\top},
-\mathrm{vec}([\tilde{X}_{i1}^{\top},\tilde{U}_{i1}^{\top}]^{\top})^{\top}]^{\top}$. Let $\vartheta_{i,0}$ and $\vartheta_{i,1}$ be certain positive constants. By Lemma \ref{L1} and $0<\rho(\mathcal{H}^{\mu})<1$, we have
$\|\kappa\bar{\Omega}_{i}^{\top}\tilde{\eta}_{i}\|\leq\vartheta_{i,1}(\rho(\mathcal{H}^{\mu}))^{t_{0}}$.
Then, from \eqref{21a}, one has
\begin{equation}\label{23}
\begin{aligned}
\|\tilde{\chi}^{n}_{i1}\|\leq&(\vartheta_{i,0})^{n}\|\tilde{\chi}^{0}_{i1}\|
+\big((\vartheta_{i,0})^{n-1}+1\big)\vartheta_{i,1}(\rho(\mathcal{H}^{\mu}))^{t_{0}},
\end{aligned}
\end{equation}
where $\vartheta_{i,0}=\rho(I-\kappa\bar{\Omega}_{i}^{\top}\bar{\Omega}_{i})$ with $0<\kappa_{i}<2/\rho(\Omega_{i}^{\top}\Omega_{i})$. We have
\begin{equation}\label{24}
\begin{aligned}
\lim_{n\rightarrow\infty}\|\tilde{\chi}^{n}_{i}\|\leq\vartheta_{i,1}(\rho(\mathcal{H}^{\mu}))^{t_{0}}.
\end{aligned}
\end{equation}
Since $\vartheta_{i,1}$ is positive constant and $\vartheta_{i,1}(\rho(\mathcal{H}^{\mu}))^{t_{0}}$ is sufficiently small if  $\rho(\mathcal{H}^{\mu})$ is sufficiently small or $t_{0}$ is sufficiently large. The solution $\hat{\chi}^{n}_{i}$ converges to the solution of \eqref{19a}. The proof is completed.$\Box$

\begin{remark}
Theorem \ref{the2} proves that a model-based approximate solution to the regulator equations \eqref{6a}-\eqref{6b} can be obtained by iteration equation \eqref{20}. This is an online iteration computational method, which is different from \cite{jiang2019optimal,jiang2023reinforcement}. We analyze the effect of the observer on the approximate solution to the output regulator equations under the constraints of the communication topology. Note that we introduce the matrix $M_{i}$, which is replaced with $A_{i}^{\top}P_{i}$ in the data-driven design phase to reconstruct the model-free $\Omega_{i}$ and $\hat{\eta}_{i}$.
\end{remark}

\subsection{Model-Based SPI Solution}
In this section, we present a method for computing the stabilizing control policy for system \eqref{8a} and Lemma \ref{L3}, which is subsequently replaced with two data-driven versions.

For any system \eqref{8a}, there must exist a positive definite unknown constant $\beta_{i}$ satisfying
\begin{equation}\label{25}
\begin{aligned}
0<\beta_{i}<\tilde{\rho}_{i}:=\frac{1}{\rho(A_{i}-B_{i}\tilde{K}_{i}^0)}
\end{aligned}
\end{equation}
such that $\rho((\tilde{\rho}_{i}-\beta_{i})(A_{i}-B_{i}\tilde{K}_{i}^0))<1$, where $\tilde{K}_{i}^0$ is the arbitrarily bounded control gain. The SPI is designed as follows.

\begin{lemma}\label{L4}({\it SPI})
 For $i=1,\ldots,N$, given arbitrarily bounded policy $\tilde{K}_{i}^0$, constants $\tilde{\beta}_{i}$ and $\alpha_{i}^{0}$ satisfying $\tilde{\beta}_{i}:=\tilde{\rho}_{i}-\beta_{i}$ and $\tilde{\rho}_{i}>\beta_{i}>\alpha_{i}^0>0$. For $k=0,1,2\ldots$, solve $\tilde{P}^{k}_{i}$ by Lyapunov equation,
\begin{equation}\label{26}
\begin{aligned}
&(\tilde{A}_{i}^{k})^{\top}\tilde{P}^{k}_{i}\tilde{A}_{i}^{k}-\tilde{P}^{k}_{i}=-Q_{i}-(\tilde{K}_{i}^{k})^{\top}R_{i}\tilde{K}_{i}^{k},
\end{aligned}
\end{equation}
where $\tilde{A}_{i}^{k}:=(\tilde{\beta}_{i}+\sum_{m=0}^{k}\alpha_{i}^{m})(A_{i}-B_{i}\tilde{K}_{i}^{k})$. Update control policy $\tilde{K}_{i}^{k+1}$ and iteration step-size $\alpha_{i}^{k+1}$ by
\begin{equation}\label{27}
\begin{aligned}
\tilde{K}_{i}^{k+1}=&(\tilde{\beta}_{i}+\sum_{m=0}^{k}\alpha_{i}^{m})^{2}\big(R_{i}+(\tilde{\beta}_{i}+\sum_{m=0}^{k}\alpha_{i}^{m})^{2}B_{i}^{\top}\tilde{P}^{k}_{i}B_{i}\big)^{-1}\\
&\times B_{i}^{\top}\tilde{P}^{k}_{i}A_{i},
\end{aligned}
\end{equation}
\begin{equation}\label{28}
\begin{aligned}
0<\alpha_{i}^{k+1}<\frac{1}{\rho(A_{i}-B_{i}\tilde{K}_{i}^{k+1})}-(\tilde{\beta}_{i}+\sum_{m=0}^{k}\alpha_{i}^{m}).
\end{aligned}
\end{equation}
Then, there are: (i). $\rho(A_{i}-B_{i}\tilde{K}_{i}^{k+1})<1/(\tilde{\beta}_{i}+\sum_{m=0}^{k+1}\alpha_{i}^{m})$; (ii). $\alpha_{i}^{k+1}$ exists and is bounded.
\end{lemma}

{\it Proof.}
Prove (i) by {\it Induction}. For $i=1,\ldots,N$ and $k=0$, one has $\rho((\tilde{\beta}_{i}+\alpha_{i}^{0})(A_{i}-B_{i}\tilde{K}_{i}^0))<\tilde{\rho}_{i}\rho(A_{i}-B_{i}\tilde{K}_{i}^0)=1$. Clearly, $\tilde{K}_{i}^0$ is the stabilizing control gain of pair $((\tilde{\beta}_{i}+\alpha_{i}^{0})A_{i},(\tilde{\beta}_{i}+\alpha_{i}^{0})B_{i})$. It follows from Lemma \ref{L3} that $\rho((\tilde{\beta}_{i}+\alpha_{i}^{0})(A_{i}-B_{i}\tilde{K}_{i}^1))<1$. There exists $\alpha_{i}^{1}$ satisfying \eqref{28} due to $1/\rho(A_{i}-B_{i}\tilde{K}_{i}^{1})-(\tilde{\beta}_{i}+\alpha_{i}^{0})>0$. Hence, one has $\rho(A_{i}-B_{i}\tilde{K}_{i}^{1})<1/(\tilde{\beta}_{i}+\sum_{m=0}^{1}\alpha_{i}^{m})$. For $i=1,\ldots,N$ and $k=0,1$, Conclusion (i) holds.

Suppose that Conclusion (i) holds for $k=z$ and $i=1,\ldots,N$, where $z\in\mathbb{N}$. There is $\rho((\tilde{\beta}_{i}+\sum_{m=0}^{z}\alpha_{i}^{m})(A_{i}-B_{i}\tilde{K}_{i}^{z}))<1$.
This implies $\tilde{K}_{i}^{z}$ is a stabilizing gain of pair $((\tilde{\beta}_{i}+\sum_{m=0}^{z}\alpha_{i}^{m})A_{i},(\tilde{\beta}_{i}+\sum_{m=0}^{z}\alpha_{i}^{m})B_{i})$. By Lemma \ref{L3}, one has $\rho((\tilde{\beta}_{i}+\sum_{m=0}^{z}\alpha_{i}^{m})(A_{i}-B_{i}\tilde{K}_{i}^{z+1}))<1$. Similarly, there exist $\alpha_{i}^{z+1}$ satisfying $0<\alpha_{i}^{z+1}<1/\rho(A_{i}-B_{i}\tilde{K}_{i}^{z+1})-(\tilde{\beta}_{i}+\sum_{m=0}^{z+1}\alpha_{i}^{m})$. Thus, we have $\rho((\tilde{\beta}_{i}+\sum_{m=0}^{z+1}\alpha_{i}^{m})(A_{i}-B_{i}\tilde{K}_{i}^{z+1}))<1$. Conclusion (i) holds for
$k=z+1$ and $i=1,\ldots,N$. Therefore, Conclusion (i) is true.

Proof of Conclusion (ii). For $i=1,\ldots,N$, since $\alpha_{i}^0$ and $\tilde{K}_{i}^0$ are bounded, $\tilde{P}_{i}^{0}$ is bounded from \eqref{26}. Hence, $\tilde{K}_{i}^1$ also is bounded from \eqref{27}. Recursively, we have that $\tilde{K}_{i}^{k+1}$ is bounded if $\alpha_{i}^{k}$ is bounded. Therefore, there is $0<1/\rho(A_{i}-B_{i}\tilde{K}_{i}^{k+1})-(\tilde{\beta}_{i}+\sum_{m=0}^{k+1}\alpha_{i}^{m})<\infty$. It follows that $\alpha_{i}^{k+1}$ exists and is bounded.
$\Box$

\begin{remark}
According to Conclusion (i) of Lemma \ref{L4}, the stabilizing control policy $\tilde{K}_{i}^{k+1}$ can be obtained if $\beta_{i}+\sum_{m=0}^{k+1}\alpha_{i}^{m}\geq1$. Let $K_{i}^0=\tilde{K}_{i}^{k+1}$ in Lemma \ref{L3}, the optimal control gain can be obtained by PI. Different from Lemma \ref{L3}, Lemma \ref{L4} can start with arbitrary initial control policy. The algorithm is equivalent to iterating the stabilizing virtual system $(\beta_{i}+\alpha_{i}^{0})(A_{i}-B_{i}\tilde{K}_{i}^{0})$ step-by-step to the actual system $(A_{i}-B_{i}\tilde{K}_{i}^{k+1})$ until $\beta_{i}+\sum_{m=0}^{k+1}\alpha_{i}^{m}\geq1$.
\end{remark}

\section{Data-Driven COOT By Off-Policy Learning} \label{section:4}
\subsection{Data Collection And Stabilizing Policy Computation}\label{sec4.1.1}
For $i=1,\ldots,N$ and $l=0,2,3,\ldots,h_{i}$, define $\tilde{x}_{i1}(t)=x_{i}(t)-\hat{X}_{i1}\zeta_{i}(t)$ and $\tilde{x}_{il}(t)=x_{i}(t)-X_{il}\zeta_{i}(t)$, where $X_{il}$ has been set in Section \ref{section:3:1}. By using \eqref{1a}, \eqref{6} and \eqref{15}, we have
\begin{equation}\label{29}
\begin{aligned}
\tilde{x}_{il}(t+1)=A_{i}\tilde{x}_{il}(t)+B_{i}u_{i}(t)+\pi_{il}\zeta_{i}(t)+\varrho_{il}(t),
\end{aligned}
\end{equation}
where $\pi_{il}=-\Upsilon_{i}(X_{il})$, $\varrho_{il}(t)=-X_{il}\tilde{E}_{i}\zeta_{i}(t)-\mu_{i}X_{il}E_{i}\sum_{j=0}^{N}a_{ij}(\zeta_{j}(t)-\zeta_{i}(t))$,
$\pi_{i1}$ and $\varrho_{i1}(t)$ are obtained by replacing $X_{i1}$ as $\hat{X}_{i1}$, respectively.
By using the definition of $\tilde{A}_{i}^{k}$, we rewrite system \eqref{29} as
\begin{equation}\label{31}
\begin{aligned}
\tilde{x}_{il}(t+1)=&(\tilde{\beta}_{i}+\sum_{m=0}^{k}\alpha_{i}^{m})^{-1}\tilde{A}_{i}^{k}\tilde{x}_{il}(t)\\
&+B_{i}(\tilde{K}_{i}^{k}\tilde{x}_{il}(t)+u_{i}(t))+\pi_{il}\zeta_{i}(t)+\varrho_{il}(t).
\end{aligned}
\end{equation}
Using \eqref{26}, \eqref{27} and \eqref{31}, there is
\begin{equation}\label{32}
\begin{aligned}
&\tilde{x}_{il}^{\top}(t+1)\tilde{P}_{i}^{k}\tilde{x}_{il}(t+1)-\tilde{x}_{il}^{\top}(t)\tilde{P}_{i}^{k}\tilde{x}_{il}(t)\\
=&(\tilde{\beta}_{i}+\sum_{m=0}^{k}\alpha_{i}^{m})^{-2}\tilde{x}_{il}^{\top}(t)(-Q_{i}-(\tilde{K}_{i}^{k})^{\top}R_{i}\tilde{K}_{i}^{k})\tilde{x}_{il}(t)\\
&+\big((\tilde{\beta}_{i}+\sum_{m=0}^{k}\alpha_{i}^{m})^{-2}-1\big)\tilde{x}_{il}^{\top}(t)\tilde{P}_{i}^{k}\tilde{x}_{il}(t)\\
&+u_{i}^{\top}(t)B_{i}^{\top}\tilde{P}_{i}^{k}B_{i}u_{i}(t)-\tilde{x}_{il}^{\top}(t)\tilde{K}_{i}^{k\top}B_{i}^{\top}\tilde{P}_{i}^{k}B_{i}\tilde{K}_{i}^{k}\tilde{x}_{il}(t)\\
&+2\tilde{x}_{il}^{\top}(t)A_{i}^{\top}\tilde{P}_{i}^{k}B_{i}(u_{i}(t)+\tilde{K}_{i}^{k}\tilde{x}_{il}(t))+\zeta_{i}^{\top}(t)\pi_{il}^{\top}\tilde{P}_{i}^{k}\pi_{il}\zeta_{i}(t)\\
&+2u_{i}^{\top}(t)B_{i}\tilde{P}_{i}^{k}\pi_{il}\zeta_{i}(t)+2\tilde{x}_{il}^{\top}(t)A_{i}^{\top}\tilde{P}_{i}^{k}\pi_{il}\zeta_{i}(t)+\omega_{il}(t),
\end{aligned}
\end{equation}
where $\omega_{il}(t)$ is caused by $\varrho_{il}(t)$ as
$\omega_{il}(t)
=2\tilde{x}_{il}^{\top}(t)(A_{i}-B_{i}\tilde{K}_{i}^{k})^{\top}\tilde{P}_{i}^{k}\varrho_{il}(t)+2\zeta_{i}^{\top}(k)\pi_{il}^{\top}\tilde{P}_{i}^{k}\varrho_{il}(t)
+2(u_{i}(t)+\tilde{K}_{i}^{k}\tilde{x}_{il}(t))^{\top}B_{i}^{\top}\tilde{P}_{i}^{k}\varrho_{il}(t)+\varrho_{il}^{\top}(t)\tilde{P}_{i}^{k}\varrho_{il}(t)$.
We can obtain the following matrices by collecting data at time interval $[t_{0}, t_{f}]$:
$\theta_{\tilde{x}_{il}}=[\mathrm{vecv}(\tilde{x}_{il}(t_{1}))-\mathrm{vecv}(\tilde{x}_{il}(t_{0})),\ldots,
\mathrm{vecv}(\tilde{x}_{il}(t_{f+1}))-\mathrm{vecv}(\tilde{x}_{il}(t_{f}))]^{\top}$,
$\Gamma_{\tilde{x}_{il}}=[\mathrm{vecv}(\tilde{x}_{il}(t_{0})),\ldots,\mathrm{vecv}(\tilde{x}_{il}(t_{f}))]^{\top}$,
$\Gamma_{\zeta_{i}}=[\mathrm{vecv}(\zeta_{i}(t_{0})),\ldots,\mathrm{vecv}(\zeta_{i}(t_{f}))]^{\top}$,
$\Gamma_{u_{i}}=[\mathrm{vecv}(u_{i}(t_{0})),\ldots,\mathrm{vecv}(u_{i}(t_{f}))]^{\top}$,
$\Gamma_{k\tilde{x}_{il}}=[\mathrm{vecv}(\tilde{K}_{i}^{k}\tilde{x}_{il}(t_{0})),\ldots,\mathrm{vecv}(\tilde{K}_{i}^{k}\tilde{x}_{il}(t_{f}))]^{\top}$,
$\Gamma_{\tilde{x}\tilde{x}_{il}}=[\tilde{x}_{il}(t_{0})\otimes\tilde{x}_{il}(t_{0}),\ldots,\tilde{x}_{il}(t_{f})\otimes\tilde{x}_{il}(t_{f})]^{\top}$,
$\Gamma_{u_{i}\tilde{x}_{il}}=[u_{i}(t_{0})\otimes\tilde{x}_{il}(t_{0}),\ldots,u_{i}(t_{f})\otimes\tilde{x}_{il}(t_{f})]^{\top}$,
$\Gamma_{\zeta_{i}\tilde{x}_{il}}=[\zeta_{i}(t_{0})\otimes\tilde{x}_{il}(t_{0}),\ldots,\zeta_{i}(t_{f})\otimes\tilde{x}_{il}(t_{f})]^{\top}$,
$\Gamma_{\zeta_{i}u_{i}}=[\zeta_{i}(t_{0})\otimes u_{i}(t_{0}),\ldots,\zeta_{i}(t_{f})\otimes u_{i}(t_{f})]^{\top}$,
$\Gamma_{\omega_{il}}=[\omega_{il}(t_{0}),\ldots,\omega_{il}(t_{f})]^{\top}$.

Similar to Theorem \ref{the2}, the data error $\Gamma_{\omega_{il}}$ is also affected by the observation errors.
Let $\vartheta_{i,b}$ be certain positive constants, where $b=2,3,4$. From Lemma \ref{L1}, one has that $\|\sum_{j=0}^{N}a_{ij}(\zeta_{j}(t_{0})-\zeta_{i}(t_{0}))\|<\vartheta_{i,2}$ and $\|\sum_{j=0}^{N}a_{ij}(E_{j}(t_{0})-E_{i}(t_{0}))\|<\vartheta_{i,3}$, hence, there exists a
small constant $\vartheta_{i,4}$ satisfying $\|\varrho_{il}(t_{0})\|<\vartheta_{i,4}$. For $\forall t>t_{0}$, $\vartheta_{i,4}$ will be sufficiently small as $\vartheta_{i,2}$ and $\vartheta_{i,3}$ are chosen to be sufficiently small. According to Lemma \ref{L1}, $\vartheta_{i,2}$ and $\vartheta_{i,3}$ can be arbitrarily small by setting a suitable $\mu_{i}$. Based on the above analysis and similar to \cite[Thm. 2]{chen2022robust}, \cite{jiang2023reinforcement}, \eqref{32} can be approximated as
\begin{equation}\label{36}
\begin{aligned}
\tilde{\phi}_{il}^{k}[&\mathrm{vecs}(\tilde{P}_{i}^{k})^{\top},\mathrm{vec}(\tilde{L}_{1,i}^{k})^{\top},\mathrm{vecs}(\tilde{L}_{2,i}^{k})^{\top},
\mathrm{vec}(\tilde{L}_{3,il}^{k})^{\top},\\
&\mathrm{vec}(\tilde{L}_{4,il}^{k})^{\top},\mathrm{vecs}(\tilde{L}_{5,il}^{k})^{\top}]^{\top}=\tilde{\psi}_{il}^{k},
\end{aligned}
\end{equation}
where $\tilde{\phi}_{il}^{k}=[\theta_{\tilde{x}_{il}}-((\tilde{\beta}_{i}+\sum_{m=0}^{k}\alpha_{i}^{m})^{-2}-1)\Gamma_{\tilde{x}_{il}},
-2\Gamma_{\tilde{x}\tilde{x}_{il}}(I_{n_{i}}\otimes\tilde{K}_{i}^{k\top})-2\Gamma_{u_{i}\tilde{x}_{il}},\Gamma_{k\tilde{x}_{il}}-\Gamma_{u_{i}},
-2\Gamma_{\zeta_{i}\tilde{x}_{il}},-2\Gamma_{\zeta_{i}u_{i}},-\Gamma_{\zeta_{i}}]$,
$\tilde{L}_{1,i}^{k}=A_{i}^{\top}\tilde{P}_{i}^{k}B_{i}$,$\tilde{L}_{2,i}^{k}=B_{i}^{\top}\tilde{P}_{i}^{k}B_{i}$, $\tilde{L}_{3,il}^{k}=A_{i}^{\top}\tilde{P}_{i}^{k}\pi_{il}$,
$\tilde{L}_{4,il}^{k}=B_{i}^{\top}\tilde{P}_{i}^{k}\pi_{il}$, $\tilde{L}_{5,il}^{k}=\pi_{il}^{\top}\tilde{P}_{i}^{k}\pi_{il}$,
$\tilde{\psi}_{il}^{k}=(\tilde{\beta}_{i}+\sum_{m=0}^{k}\alpha_{i}^{m})^{-2}\Gamma_{\tilde{x}_{il}}\mathrm{vecs}(-\bar{Q}_{i}^{k})$,
$\bar{Q}_{i}^{k}:=Q_{i}+(\tilde{K}_{i}^{k})^{\top}R_{i}\tilde{K}_{i}^{k}$, $k=0,1,\ldots$ and $l=0,1,\ldots,h_{i}$.

\begin{lemma}\label{L5}
 For $i=1,\ldots,N$, if the condition
\begin{equation}\label{37}
\begin{aligned}
\mathrm{rank}&([\Gamma_{\tilde{x}\tilde{x}_{il}},\Gamma_{u_{i}},\Gamma_{\zeta_{i}},\Gamma_{u_{i}\tilde{x}_{il}},\Gamma_{\zeta_{i}\tilde{x}_{il}},\Gamma_{\zeta_{i}u_{i}}])\\
&=(n_{i}+m_{i}+n_{v})(n_{i}+m_{i}+n_{v}+1)/2
\end{aligned}
\end{equation}
is satisfied, $\tilde{\phi}_{il}^{k}$ has full column-rank.
\end{lemma}

The full column-rank condition similar to Lemma \ref{L5} is commonly found in data-driven RL methods such as \cite{jiang2024adaptive,chen2022robust,li2024data}. By Lemma \ref{L5}, for $k=0,1,2,\ldots$, \eqref{36} can be solved by
\begin{equation}\label{38}
\begin{aligned}
&[\mathrm{vecs}(\tilde{P}_{i}^{k})^{\top},\mathrm{vec}(\tilde{L}_{1,i}^{k})^{\top},\mathrm{vecs}(\tilde{L}_{2,i}^{k})^{\top},
\mathrm{vec}(\tilde{L}_{3,il}^{k})^{\top},\\
&\mathrm{vec}(\tilde{L}_{4,il}^{k})^{\top},\mathrm{vecs}(\tilde{L}_{5,il}^{k})^{\top}]^{\top}=(\tilde{\phi}_{il}^{k\top}\tilde{\phi}_{il}^{k})^{-1}\tilde{\phi}_{il}^{k\top}\tilde{\psi}_{il}^{k}.
\end{aligned}
\end{equation}
According to \eqref{27}, update the control gain by
\begin{align}
\tilde{K}_{i}^{k+1}=&(\tilde{\beta}_{i}+\sum_{m=0}^{k}\alpha_{i}^{m})^{2}\big(R_{i}+(\tilde{\beta}_{i}
+\sum_{m=0}^{k}\alpha_{i}^{m})^{2}\tilde{L}_{2,i}^{k}\big)^{-1}\tilde{L}_{1,i}^{k\top}.\label{39}
\end{align}

\subsection{Determine $\tilde{\beta}_{i}+\alpha_{i}^{0}$ and choose $\alpha_{i}^{k+1}$ by model-free methods}

{\bf Initialize} a sufficiently small constant $\alpha_{i}^{0}>0$ and a monotonic sequence $\{\tilde{\beta}_{i}^z\}$ satisfying $0<\tilde{\beta}_{i}^{z+1}<\tilde{\beta}_{i}^{z}<1$ and $\lim_{z\rightarrow\infty}\tilde{\beta}_{i}^{z}=0$ for $\forall z\in\mathbb{N}$. Set $z\leftarrow0$ and $\tilde{\beta}_{i}\leftarrow\tilde{\beta}_{i}^0$, {\bf repeat:}  1). {\bf if} $\tilde{P}_{i}^{0}\leq0$ from \eqref{38}, {\bf then} $z\leftarrow z+1$, $\tilde{\beta}_{i}\leftarrow\tilde{\beta}_{i}^z$ and calculate $\tilde{P}_{i}^{0}$ by \eqref{38}; {\bf else end}; 2). {\bf output} $\tilde{\beta}_{i}\leftarrow\tilde{\beta}_{i}^z$ and $\alpha_{i}^{0}$.
\begin{remark}
If \eqref{37} is satisfied, solving $\tilde{P}_{i}^{k}$ by \eqref{38} is equivalent to solving $\tilde{P}_{i}^{k}$ by \eqref{26}. For $k = 0$, if $\tilde{P}_{i}^{0}>0$, it can be deduced from Lyapunnov theory that $(\tilde{\beta}_{i}+\alpha_{i}^{0})(A_{i}-B_{i}\tilde{K}_{i}^0)$ is Schur. Therefore, we can use the above iterations to find the coefficient $\tilde{\beta}_{i}+\alpha_{i}^{0}$ such that the system $((\tilde{\beta}_{i}+\alpha_{i}^{0})A_{i},(\tilde{\beta}_{i}+\alpha_{i}^{0})B_{i})$ is stabilized by any bounded initial control gain $\tilde{K}_{i}^{0}$. 
\end{remark}

{\bf Scheme 1.} We design the following selection scheme for $\alpha_{i}^{k+1}$ as
\begin{equation}\label{44}
\begin{aligned}
\alpha_{i}^{k+1}=a_{i}(\tilde{\beta}_{i}+\sum_{m=0}^{k}\alpha_{i}^{m})\bigg(\sqrt{\frac{\sigma_{\min}(\bar{Q}_{i}^{k+1})}{\sigma_{\max}(\tilde{P}_{i}^{k}-\bar{Q}_{i}^{k+1})}+1}-1\bigg),
\end{aligned}
\end{equation}
where $0<a_{i}\leq1$, $\tilde{P}_{i}^{k}$ and $\bar{Q}_{i}^{k+1}$ are obtained by \eqref{38}-\eqref{39}.

\begin{theorem}\label{the3}
For $i=1,\ldots,N$, given the control gain $\tilde{K}_{i}^{k+1}$ by \eqref{27}. If $\alpha_{i}^{k+1}$ is updated by Scheme 1, then system $\tilde{A}^{k+1}_{i}$ is Schur.
\end{theorem}

{\it Proof.}
For $i=1,\ldots,N$, since $\tilde{K}_{i}^{k+1}$ from \eqref{27} is the stabilizing
control gain of $(\tilde{\beta}_{i}+\sum_{m=0}^{k}\alpha_{i}^{m})(A_{i}-B_{i}\tilde{K}_{i}^{k+1})$. There must exist the unique positive solution to $[(\tilde{\beta}_{i}+\sum_{m=0}^{k}\alpha_{i}^{m})(A_{i}-B_{i}\tilde{K}_{i}^{k+1})]^{\top}\hat{\tilde{P}}_{i}^{k+1}[(\tilde{\beta}_{i}+\sum_{m=0}^{k}\alpha_{i}^{m})(A_{i}
-B_{i}\tilde{K}_{i}^{k+1})]-\hat{\tilde{P}}_{i}^{k+1}=-\bar{Q}^{k+1}_{i}$.
Viewing $(\tilde{\beta}_{i}+\sum_{m=0}^{k}\alpha_{i}^{m})A_{i}$ and $(\tilde{\beta}_{i}+\sum_{m=0}^{k}\alpha_{i}^{m})B_{i}$ as $A_{i}$ and $B_{i}$ respectively in Lemma \ref{L3}, we get $0<\hat{\tilde{P}}_{i}^{k+1}\leq\tilde{P}_{i}^{k}$ by Conclusion (ii) in Lemma \ref{L3}. Then, one has
\begin{equation}\label{46}
\begin{aligned}
&(\tilde{A}^{k+1}_{i})^{\top}\hat{\tilde{P}}_{i}^{k+1}\tilde{A}^{k+1}_{i}-\hat{\tilde{P}}_{i}^{k+1}\\
\overset{(a)}{=}&\big((\frac{\tilde{\beta}_{i}+\sum_{m=0}^{k+1}\alpha_{i}^{m}}{\tilde{\beta}_{i}+\sum_{m=0}^{k}\alpha_{i}^{m}})^{2}-1\big)(\hat{\tilde{P}}_{i}^{k+1}-\bar{Q}^{k+1}_{i})-\bar{Q}^{k+1}_{i}\\
\overset{(b_{1})}{\leq}&\big((1+\frac{\alpha_{i}^{k+1}}{\tilde{\beta}_{i}+\sum_{m=0}^{k}\alpha_{i}^{m}})^{2}-1\big)(\tilde{P}_{i}^{k}-\bar{Q}^{k+1}_{i})-\bar{Q}^{k+1}_{i}\\
\overset{(c)}{<}&\sigma_{\min}(\bar{Q}^{k+1}_{i})I_{n_{i}}-\bar{Q}^{k+1}_{i}<0,
\end{aligned}
\end{equation}
where $(b_{1})$ is obtained by $0<\hat{\tilde{P}}_{i}^{k+1}\leq\tilde{P}_{i}^{k}$ and $(c)$ is obtained by \eqref{44}. Since the right hand side of Eq. $(a)$ in \eqref{46} is a monotonically increasing function on the interval \eqref{44} with respect to $\alpha_{i}^{k+1}$, any choice of $\alpha_{i}^{k+1}$ on interval \eqref{44} ensures the stability of system $\tilde{A}^{k+1}_{i}$. Since $\tilde{A}^{k+1}_{i}$ is Schur, then one gets \eqref{28}.
$\Box$

\subsection{SPI-Based Data-Driven COOT Algorithm 1}
 If $\beta_{i}+\sum_{m=0}^{k+1}\alpha_{i}^{m}\geq1$ is reached, then let $K^j_{i}=\tilde{K}_{i}^{k+1}$ be the stabilizing control gain with $j=0$. Similar to \eqref{36} and from Lemma \ref{L3}, if \eqref{37} is satisfied, $i=1,\ldots,N$ and $j = 0,1,2,\ldots$, we have
\begin{equation}\label{49}
\begin{aligned}
&[\mathrm{vecs}({P}_{i}^{j})^{\top},\mathrm{vec}({L}_{1,i}^{j})^{\top},\mathrm{vecs}({L}_{2,i}^{j})^{\top},\mathrm{vec}({L}_{3,il}^{j})^{\top},\\
&\mathrm{vec}({L}_{4,il}^{j})^{\top},\mathrm{vecs}({L}_{5,il}^{j})^{\top}]^{\top}=({\phi}_{il}^{j\top}{\phi}_{il}^{j})^{-1}{\phi}_{il}^{j\top}{\psi}_{il}^{j},
\end{aligned}
\end{equation}
where
${\phi}_{il}^{j}=[\theta_{\tilde{x}_{il}},
-2\Gamma_{\tilde{x}\tilde{x}_{il}}(I_{n_{i}}\otimes{K}_{i}^{j\top})-2\Gamma_{u_{i}\tilde{x}_{il}},\Gamma_{j\tilde{x}_{il}}
-\Gamma_{u_{i}},-2\Gamma_{\zeta_{i}\tilde{x}_{il}},-2\Gamma_{\zeta_{i}u_{i}},-\Gamma_{\zeta_{i}}]$,
${L}_{1,i}^{j}=A_{i}^{\top}{P}_{i}^{j}B_{i}$, ${L}_{2,i}^{j}=B_{i}^{\top}{P}_{i}^{j}B_{i}$, ${L}_{3,il}^{j}=A_{i}^{\top}{P}_{i}^{j}\pi_{il}$,
${L}_{4,il}^{j}=B_{i}^{\top}{P}_{i}^{j}\pi_{il}$, ${L}_{5,il}^{j}=\pi_{il}^{\top}{P}_{i}^{j}\pi_{il}$,
${\psi}_{il}^{j}=\Gamma_{\tilde{x}\tilde{x}_{il}}\mathrm{vecs}(-Q_{i}-{K}_{i}^{j\top}R_{i}{K}_{i}^{j})$,
$\Gamma_{j\tilde{x}_{il}}=[\mathrm{vecv}({K}_{i}^{j}\tilde{x}_{il}(t_{0})),\ldots,\mathrm{vecv}({K}_{i}^{j}\tilde{x}_{il}(t_{f}))]^{\top}$.
Update control gain by
\begin{equation}\label{50}
\begin{aligned}
{K}_{i}^{j+1}=(R_{i}+{L}_{2,i}^{j})^{-1}{L}_{1,i}^{j\top}.
\end{aligned}
\end{equation}

Based on \eqref{49}, let $M_{i}=A_{i}^{\top}P_{i}^{j}$ in \eqref{19b}. Note that $\pi_{il}=-\Upsilon_{i}(X_{il})$. Since $\Upsilon_{i}(X_{il})=X_{il}E-A_{i}X_{il}=-\pi_{il}$, there is $\bar{\Upsilon}_{i}(X_{il},U_{il})=-\pi_{il}-B_{i}U_{il}$. Therefore, we have $M_{i}\bar{\Upsilon}_{i}(X_{il},U_{il})=-A_{i}^{\top}P_{i}^{j}\pi_{il}-A_{i}^{\top}P_{i}^{j}B_{i}U_{il}
=-{L}_{3,il}^{j}-L_{1,i}^{j}U_{il}$. The $\Omega_{i}$ and $\hat{\eta}_{i}$ can be obtained by replacing $M_{i}\bar{\Upsilon}_{i}(X_{il},U_{il})$ in \eqref{19b}.
\begin{algorithm}[!h]
	\caption{Lyapunov-Based Off-Policy SPI Algorithm}\label{alg1}
	\begin{algorithmic}[1]
		\STATE {\bf{Initialize:}}  Select arbitrary control policy $\tilde{K}_{i}^{0}$, sufficiently small positive constants $\varepsilon_{i,1}$, $\varepsilon_{i,2}$ and $\alpha_{i}^{0}$, a monotonically decreasing sequence $\{\tilde{\beta}_{i}^{z}\}$, where $0<\tilde{\beta}_{i}^{z}<1$ for $\forall z$.
$k\leftarrow0$, $z\leftarrow0$, $l\leftarrow0$, $j\leftarrow0$, $\tilde{\beta}_{i}\leftarrow\tilde{\beta}_{i}^{0}$.
       \STATE{\bf Pre-collection:} Set the appropriate $\mu_{i}$. Obtain $(\hat{X}_{i1},\hat{U}_{i1})$ by \eqref{16}.
       \STATE{\bf Data-collection:} Collect data on the time interval $[t_{0}, t_{f}]$ until \eqref{37} is satisfied.\\
      \STATE{\bf Online iteration:} {\it (Steps 5-13)}
      \STATE{ If $\tilde{P}_{i}^{0}\leq0$ from \eqref{38}, then $z\leftarrow z+1$, $\tilde{\beta}_{i}\leftarrow\tilde{\beta}_{i}^z$ and calculate $\tilde{P}_{i}^{0}$ by \eqref{38}, otherwise output $\tilde{\beta}_{i}\leftarrow\tilde{\beta}_{i}^z$ and go to the next step;}
      \STATE{Solve $\tilde{P}_{i}^{k}$, $\tilde{L}_{1,i}^{k}$ and $\tilde{L}_{2,i}^{k}$ by \eqref{38}, update $\tilde{K}_{i}^{k+1}$ by \eqref{39};}
      \STATE{Choose $\alpha_{i}^{k+1}$ by Scheme 1. If $\tilde{\beta}_{i}+\sum_{m=0}^{k+1}\alpha_{i}^{m}\geq1$, $K_{i}^0\leftarrow\tilde{K}_{i}^{k+1}$ and go to the next step; Otherwise $k\leftarrow k+1$ and go to Step 6;}
      \STATE{Solve ${P}_{i}^{j}$, ${L}_{1,i}^{j}$ and ${L}_{2,i}^{j}$ by \eqref{49}, update policy $K_{i}^{j+1}$ by \eqref{50}; }
      \STATE {If $\|{P}_{i}^{j}-{P}_{i}^{j-1}\|\leq\varepsilon_{i,1}$, go to the next step; Otherwise $j\leftarrow j+1$ and go to Step 8;}
      \STATE {Get $\Omega_{i}$ and $\hat{\eta}_{i}$ by $M_{i}\bar{\Upsilon}_{i}(X_{il},U_{il})=-{L}_{3,il}^{j}-L_{1,i}^{j}U_{il}$;}
      \STATE {Choose parameter $0<\kappa_{i}<2/\rho(\Omega_{i}^{\top}\Omega_{i})$;}
      \STATE {Calculate $\hat{\chi}_{i}^{n}$ by \eqref{20}; }
      \STATE {If $\|\hat{\chi}_{i}^{n+1}-\hat{\chi}_{i}^{n}\|\leq\varepsilon_{i,2}$, output $\hat{X}_{i}$ and $\hat{U}_{i}$ and go to the next step; Otherwise $n\leftarrow n+1$ and go to Step 12.}
      \STATE {\bf Optimal control phase:} Set $K_{i}\leftarrow K_{i}^{j}$ and ${T}_{i}\leftarrow \hat{U}_{i}+K^{j}_{i}\hat{X}_{i}$ and use the optimal controller \eqref{4}.
	\end{algorithmic}
\end{algorithm}

\begin{theorem}\label{the4}
Under Assumptions \ref{A1}-\ref{A4}, consider the distributed observer \eqref{3} and system \eqref{1a}-\eqref{2}. For $i=1,\ldots,N$, use the adaptive feedforward-feedback controller \eqref{4} with ${K}_{i}^{j}$ and ${T}_{i}$ being the optimal gains learned by Algorithm \ref{alg1}. Then, the COOT problem is solved.
\end{theorem}

 {\it Proof.}
 Based on Lemma \ref{L1} and Theorem \ref{the2}, if $t_{0}$ is sufficiently large or $\rho(\mathcal{H}^{\mu})$ sufficiently small, $\hat{\chi}_{i}^{n}$ converges to the solution of \eqref{19a}, thus the solution $(X_{i},U_{i})$ to \eqref{6} is obtained.  Similar to \cite[Thm. 2]{chen2022robust}, \cite{jiang2023reinforcement}, if $\varrho_{il}\rightarrow0$, \eqref{32} can be approximated by \eqref{36}. If \eqref{37} is satisfied, then \eqref{36} can be uniquely solved by \eqref{38}. Then, \eqref{38} is equivalent to \eqref{26}. This implies that \eqref{39} is equivalent to \eqref{27}. By Theorem \ref{the3}, updating $\alpha_{i}^{k+1}$ by using Scheme 1 satisfy \eqref{28}. Thus, steps 5-7 of Algorithm \ref{alg1} are equivalent to Lemma \ref{L4}. If $\tilde{\beta}_{i}+\sum_{m=0}^{k+1}\alpha_{i}^{m}\geq1$ is satisfied, then $\tilde{K}_{i}^{k+1}$ is a stabilizing control gain. Letting $K_{i}^0 = \tilde{K}_{i}^{k+1}$ satisfies the initial condition of Lemma \ref{L3}. If $\varrho_{il}\rightarrow0$ and \eqref{37} is satisfied, \eqref{49} is equivalent to \eqref{13}, then $K_{i}^{j+1}$ from \eqref{50} converges to the optimal control gain $K_{i}^{*}$ and $P_{i}^{j}$ converges to the optimal solution $P_{i}^*$. Therefore, the gains ${K}_{i}^j$ and $T_{i}$ learned by Algorithm \ref{alg1} converge to the optimum, respectively. Since $A_{i}-B_{i}{K}_{i}^j$ is Schur and $T_{i}$ satisfies \eqref{5}-\eqref{6}, by Theorem 1 one gets $\lim_{t\rightarrow\infty}e_{i}(t)=0$. Thus, the COOT problem is solved.
$\Box$

\begin{remark}
Since the leader state $v$ is only locally available and computing $(\hat{X}_{i1},\hat{U}_{i1})$ depends on \eqref{3} with $t_{0}$, the distributed observer \eqref{3} and its convergence are critical. The estimated state $\zeta_{i}$ is not only used for feedforward in the controller \eqref{4} but also collected for each $l$ to solve the ${L}_{3,il}^{j}$ in $M_{i}\bar{\Upsilon}_{i}$, thereby computing $(X_{i}, U_{i})$.
\end{remark}

\section{$Q$-Learning Algorithm Based on SPI}\label{section:5}
\subsection{$Q$-Learning Establishment And Stabilizing Policy Computation}\label{sec4.2.1}
For $i=1,\ldots,N$ and $l=0,2,3,\ldots,h_{i}$, define $\tilde{u}_{il}(t)=u_{i}(t)-U_{il}\zeta_{i}(t)$ and $\tilde{u}_{i1}(t)=u_{i}(t)-\hat{U}_{i1}\zeta_{i}(t)$, where $U_{il}$ has been set in Section \ref{section:3:1}.
From \eqref{29} and $\varrho_{il}(t)\rightarrow0$, we construct
\begin{equation}\label{53}
\begin{aligned}
\tilde{\xi}_{il}(t+1)=(\tilde{\beta}_{i}+\sum_{m=0}^{k}\alpha_{i}^{m})\tilde{x}_{il}(t+1),
\end{aligned}
\end{equation}
where $\tilde{x}_{il}(t+1)=A_{i}\tilde{x}_{il}(t)+B_{i}\tilde{u}_{il}(t)$ and $\gamma_{i}^{k}:=\tilde{\beta}_{i}+\sum_{m=0}^{k}\alpha_{i}^{m}$.
The value functions associated with system \eqref{53} are
\begin{subequations}\label{54}
\begin{align}
\tilde{V}_{il}(\tilde{x}_{il}(t))&=\tilde{x}_{il}^{\top}(t)\tilde{P}_{i}\tilde{x}_{il}(t),\\
\tilde{V}_{il}(\tilde{\xi}_{il}(t+1))&=\tilde{\xi}_{il}^{\top}(t+1)\tilde{P}_{i}\tilde{\xi}_{il}(t+1),
\end{align}
\end{subequations}
where $\tilde{P}_{i}=\tilde{P}_{i}^{\top}>0$.
Then, the following Bellman equation can be obtained as
$\tilde{V}_{il}(\tilde{x}_{il}(t))=\tilde{x}_{il}^{\top}(t)Q_{i}\tilde{x}_{il}(t)+\tilde{u}_{il}^{\top}(t)R_{i}\tilde{u}_{il}(t)+\tilde{V}_{il}(\tilde{\xi}_{il}(t+1))$.
Define the $Q$-function as $\tilde{Q}_{il}(\tilde{x}_{il}(t),\tilde{u}_{il}(t))=\tilde{V}_{il}(\tilde{x}_{il}(t))$. Then, one has
$\tilde{Q}_{il}(\tilde{x}_{il}(t),\tilde{u}_{il}(t)) =\tilde{x}_{il}^{\top}(t)Q_{i}\tilde{x}_{il}(t)+\tilde{u}_{il}^{\top}(t)R_{i}\tilde{u}_{il}(t)
+\tilde{V}_{il}(\tilde{\xi}_{il}(t+1))$.
By defining $Z_{il}(t)=[\tilde{x}_{il}^{\top}(t),\tilde{u}^{\top}_{il}(t)]^{\top}$, we can obtain
\begin{equation}\label{57}
\begin{aligned}
\tilde{Q}_{il}(\tilde{x}_{il}(t),\tilde{u}_{il}(t)) =&\tilde{x}_{il}^{\top}(t)Q_{i}\tilde{x}_{il}(t)+\tilde{u}_{il}^{\top}(t)R_{i}\tilde{u}_{il}(t)\\
&+(\gamma_{i}^{k})^{2}\tilde{x}_{il}^{\top}(t+1)\tilde{P}_{i}\tilde{x}_{il}(t+1)\\
=&Z_{il}^{\top}(t)\underbrace{\left[
\begin{matrix}
\tilde{H}_{i,11}&\tilde{H}_{i,12}\\
\tilde{H}_{i,21}&\tilde{H}_{i,22}\\
\end{matrix}
\right]}_{:=\tilde{H}_{i}}Z_{il}(t),
\end{aligned}
\end{equation}
where $\tilde{H}^{\top}_{i}=\tilde{H}_{i}$, $\tilde{H}_{i,11}=(\gamma_{i}^{k})^{2}A_{i}^{\top}\tilde{P}_{i}A_{i}+Q_{i}$, $\tilde{H}_{i,12}=(\gamma_{i}^{k})^{2}A_{i}^{\top}\tilde{P}_{i}B_{i}$, $\tilde{H}_{i,22}=(\gamma_{i}^{k})^{2}B_{i}^{\top}\tilde{P}_{i}B_{i}+R_{i}$.
The Bellman equation is rewritten as
\begin{equation}\label{58}
\begin{aligned}
{Z}_{il}^{\top}(t)\tilde{H}_{i}{Z}_{il}(t) =&\tilde{x}_{il}^{\top}(t)Q_{i}\tilde{x}_{il}(t)+\tilde{u}_{il}^{\top}(t)R_{i}\tilde{u}_{il}(t)\\
&+{Z}_{il}^{\top}(t+1)\tilde{H}_{i}{Z}_{il}(t+1).
\end{aligned}
\end{equation}
According to \cite{lopez2023efficient,li2024data}, the optimal control under $(X_{il},U_{il})$ is derived by $\partial\tilde{Q}_{il}(\tilde{x}_{il}(t),\tilde{u}_{il}(t))/\partial \tilde{u}_{il}=0$. Then, we have
\begin{equation}\label{59}
\begin{aligned}
\tilde{u}_{il}(t)=-(\tilde{H}_{i,22})^{-1}\tilde{H}_{i,21}\tilde{x}_{il}(t)=-\tilde{K}_{i}\tilde{x}_{il}(t),
\end{aligned}
\end{equation}
where $\tilde{H}_{i,21}=\tilde{H}_{i,12}^{\top}$.
Clearly, solving $\tilde{K}_{i}$ and $\tilde{P}_{i}$ is independent of the $l$. That is, we can obtain $\tilde{K}_{i}$ and $\tilde{H}_{i}$ by being in the  case $(X_{i0}, U_{i0})$. For $l=0$, using $\tilde{u}_{i0}(t+1)=-\tilde{K}_{i}\tilde{x}_{i0}(t+1)$ in \eqref{59}, we establish the simplified $Q$-function as
\begin{equation}\label{60}
\begin{aligned}
{Z}_{i0}^{\top}(t)\tilde{{H}}_{i}&{Z}_{i0}(t)={Z}_{i0}^{\top}(t)Q_{Ri}{Z}_{i0}(t)\\
&+(\gamma_{i}^{k})^{2}\left[
\begin{matrix}
\tilde{x}_{i0}(t+1)\\
\tilde{K}_{i}\tilde{x}_{i0}(t+1)
\end{matrix}
\right]^{\top}\tilde{{H}}_{i}\left[
\begin{matrix}
\tilde{x}_{i0}(t+1)\\
\tilde{K}_{i}\tilde{x}_{i0}(t+1)
\end{matrix}
\right],
\end{aligned}
\end{equation}
where $Q_{Ri}=\mathrm{diag}\{Q_{i},R_{i}\}$. By defining $\Pi_{i}=\gamma_{i}^{k}[I_{n_{i}},-\tilde{K}_{i}^{\top}]^{\top}[A_{i}\quad B_{i}]$ and using $\tilde{x}_{i0}(t+1)=A_{i}\tilde{x}_{i0}(t)+B_{i}\tilde{u}_{i0}(t)$, one has that \eqref{60} is equivalent to the Lyapunov equation,
\begin{equation}\label{61}
\begin{aligned}
\tilde{{H}}_{i} &=Q_{Ri}+\Pi_{i}^{\top}\tilde{{H}}_{i}\Pi_{i}.
\end{aligned}
\end{equation}
\begin{lemma} \label{L6}
For $i=1,\ldots,N$, system $\gamma_{i}^{k}(A_{i}-B_{i}\tilde{K}_{i})$ with $\gamma_{i}^{k}>0$ is stable if and only if there exists a unique positive definite solution $\tilde{{H}}_{i}$ to Lyapunov equation \eqref{61}.
\end{lemma}

{\it Proof.}
This proof can be obtained by expanding \cite[Lem. 2 and Cor. 1]{lopez2023efficient} to system $\gamma_{i}^{k}(A_{i}-B_{i}\tilde{K}_{i})$ with $\gamma_{i}^{k}>0$.
$\Box$

\begin{lemma}\label{L6a}({\it PI-based $Q$-Learning})
Given a initial stabilizing policy $K_{i}^0$ such that $\rho(A_{i}-B_{i}K_{i}^0)<1$. For $j=0,1,\ldots$ and $i=1,\ldots,N$, solve ${H}_{i}^{j}$ by
\begin{equation}\label{68}
\begin{aligned}
{Z}_{i0}^{\top}(t){H}_{i}^{j}{Z}_{i0}(t)=&{Z}_{i0}^{\top}(t)Q_{Ri}{Z}_{i0}(t)+\\
\left[
\begin{matrix}
\tilde{x}_{i0}(t+1)\\
-{K}^{j}_{i}\tilde{x}_{i0}(t+1)
\end{matrix}
\right]^{\top}&\underbrace{\left[
\begin{matrix}
{H}^{j}_{i,11}&{H}^{j}_{i,12}\\
{H}^{j}_{i,21}&{H}^{j}_{i,22}\\
\end{matrix}
\right]}_{:={H}^{j}_{i}}\left[
\begin{matrix}
\tilde{x}_{i0}(t+1)\\
-{K}^{j}_{i}\tilde{x}_{i0}(t+1)
\end{matrix}
\right],
\end{aligned}
\end{equation}
where ${H}^{j\top}_{i}={H}^{j}_{i}$, ${H}^{j}_{i,11}=A_{i}^{\top}{P}^{j}_{i}A_{i}+Q_{i}$, ${H}^{j}_{i,12}=A_{i}^{\top}\tilde{P}^{j}_{i}B_{i}$, ${H}^{j}_{i,22}=B_{i}^{\top}{P}^{j}_{i}B_{i}+R_{i}$.
Update control policy by
\begin{equation}\label{69}
\begin{aligned}
{K}_{i}^{j+1}=({H}_{i,22}^{j})^{-1}{H}_{i,21}^{j}.
\end{aligned}
\end{equation}
Then, there are: (i). $\rho(A_{i}-B_{i}K_{i}^{j+1})<1$; (ii). ${H}_{i}^{*}\leq{H}_{i}^{j+1}\leq{H}_{i}^{j}$, $\lim_{j\rightarrow\infty}{H}_{i}^{j}={H}_{i}^{*}$ and $\lim_{j\rightarrow\infty}K_{i}^{j}=K_{i}^{*}$; (iii). ${P}_{i}^{j}=[I_{n_{i}},(-{K}_{i}^{j})^{\top}]{{H}}^{j}_{i}[I_{n_{i}},(-{K}_{i}^{j})^{\top}]^{\top}$.
\end{lemma}

{\it Proof.}
The proofs of Conclusions (i) and (ii) are similar to \cite[Thms. 3-4]{lopez2023efficient}, respectively. Then, substituting the policy $\tilde{u}_{i0}(t)=-{K}^{j}_{i}\tilde{x}_{i0}(t)$ into the $Q$-function $\tilde{Q}_{il}(\tilde{x}_{il}(t),\tilde{u}_{il}(t)) =\tilde{x}_{il}^{\top}(t)Q_{i}\tilde{x}_{il}(t)+\tilde{u}_{il}^{\top}(t)R_{i}\tilde{u}_{il}(t)+\tilde{x}_{il}^{\top}(t+1){P}^{j}_{i}\tilde{x}_{il}(t+1)$ associated with \eqref{68} yields $Q_{i}(\tilde{x}_{i0}(t),-{K}^{j}_{i}\tilde{x}_{i0}(t))=\tilde{x}_{i0}^{\top}(t)(Q_{i}+K_{i}^{j\top}R_{i}K_{i}^{j})\tilde{x}_{i0}(t)
+\tilde{x}_{i0}^{\top}(t)(A_{i}-B_{i}K_{i}^{j})^{\top}P_{i}^{j}(A_{i}-B_{i}K_{i}^{j})\tilde{x}_{i0}(t)=[\tilde{x}_{i0}^{\top},(-{K}_{i}^{j}\tilde{x}_{i0})^{\top}]{{H}}^{j}_{i}[\tilde{x}_{i0}^{\top},(-{K}_{i}^{j}\tilde{x}_{i0})^{\top}]^{\top}$. Considering \eqref{13}, we have $Q_{i}(\tilde{x}_{i0}(t),-{K}^{j}_{i}\tilde{x}_{i0}(t))=\tilde{x}_{i0}^{\top}(t)P_{i}^{j}\tilde{x}_{i0}(t)$.
That is, ${P}_{i}^{j}=[I_{n_{i}},(-{K}_{i}^{j})^{\top}]{{H}}^{j}_{i}[I_{n_{i}},(-{K}_{i}^{j})^{\top}]^{\top}$.
$\Box$

Based on \eqref{59}, \eqref{60}, Lemmas \ref{L6} and \ref{L6a}, the following stabilizing $Q$-learning is established.
\begin{lemma}\label{L7}({\it SPI-based $Q$-Learning})
Given arbitrary $\tilde{K}_{i}^0$, positive constants $\tilde{\beta}_{i}$ and $\alpha_{i}^0$ satisfying $\tilde{\rho}_{i}>\beta_{i}>\alpha_{i}^0>0$. For $k=0,1,\ldots$ and $i=1,\ldots,N$, solve $\tilde{{H}}_{i}^{k}$ by
\begin{equation}\label{62}
\begin{aligned}
{Z}_{i0}^{\top}&(t)\tilde{{H}}_{i}^{k}{Z}_{i0}(t)={Z}_{i0}^{\top}(t)Q_{Ri}{Z}_{i0}(t)\\
&+(\gamma_{i}^{k})^{2}\left[
\begin{matrix}
\tilde{x}_{i0}(t+1)\\
-\tilde{K}^{k}_{i}\tilde{x}_{i0}(t+1)
\end{matrix}
\right]^{\top}\tilde{{H}}_{i}^{k}\left[
\begin{matrix}
\tilde{x}_{i0}(t+1)\\
-\tilde{K}^{k}_{i}\tilde{x}_{i0}(t+1)
\end{matrix}
\right],
\end{aligned}
\end{equation}
where $\gamma_{i}^{k}=\tilde{\beta}_{i}+\sum_{m=0}^{k}\alpha_{i}^{m}$. Update control policy by
\begin{equation}\label{63}
\begin{aligned}
\tilde{K}_{i}^{k+1}=(\tilde{H}_{i,22}^{k})^{-1}\tilde{H}_{i,21}^{k}.
\end{aligned}
\end{equation}
Choose $\alpha_{i}^{k+1}$ by \eqref{28} and update $\tilde{P}_{i}^{k}$ by
\begin{equation}\label{64}
\begin{aligned}
\tilde{P}_{i}^{k}=[I_{n_{i}},(-\tilde{K}_{i}^{k})^{\top}]\tilde{{H}}^{k}_{i}[I_{n_{i}},(-\tilde{K}_{i}^{k})^{\top}]^{\top}.
\end{aligned}
\end{equation}
Then, (i). $\rho(A_{i}-B_{i}\tilde{K}_{i}^{k+1})<1/(\tilde{\beta}_{i}+\sum_{m=0}^{k+1}\alpha_{i}^{m})$; (ii). $\alpha_{i}^{k+1}$ exists and is bounded; (iii). Lemma \ref{L7} is equivalent to Lemma \ref{L4}.
\end{lemma}

{\it Proof.}
Solving \eqref{62} is equivalent to solving $\tilde{H}^{k}_{i} =Q_{Ri}+\Pi_{i}^{k\top}\tilde{{H}}_{i}^{k}\Pi_{i}^{k}$, where $\Pi_{i}^{k}=\gamma^{k}_{i}[I_{n_{i}},(-\tilde{K}_{i}^{k})^{\top}]^{\top}[A_{i}, B_{i}]$. Conclusions (i) and (ii) of Lemma \ref{L7} can be proved by Conclusions (i) and (ii) of Lemma \ref{L6a} with the stability Lemma \ref{L6}. The detailed proof is similar to {\it Induction} in Lemma \ref{L4}, which is omitted here.

By using \eqref{64} and $[I_{n_{i}},(-\tilde{K}_{i}^{k})^{\top}]\tilde{H}^{k}_{i}[I_{n_{i}},(-\tilde{K}_{i}^{k})^{\top}]^{\top}=[I_{n_{i}},(-\tilde{K}_{i}^{k})^{\top}](Q_{Ri}
+\Pi_{i}^{k\top}\tilde{{H}}_{i}^{k}\Pi_{i}^{k})[I_{n_{i}},(-\tilde{K}_{i}^{k})^{\top}]^{\top}$, we can directly obtain Lyapunov equation \eqref{26}.
Since \eqref{63} is equivalent to \eqref{27}, then Lemma \ref{L7} is equivalent to Lemma \ref{L4}. By the equivalence, Conclusions (i) and (ii) can also be proved.$\Box$

By defining $z_{i0}^{k}(t+1)=[\tilde{x}_{i0}(t+1)^{\top},(\tilde{K}_{i}^{k}\tilde{x}_{i0}(t+1))^{\top}]^{\top}$ and collecting data on $[t_{0},t_{f}]$, one obtains the matrices as $\Gamma_{{Z}_{i0}}=[\mathrm{vecv}({Z}_{i0}(t_{0})),\mathrm{vecv}({Z}_{i0}(t_{1})),\ldots,\mathrm{vecv}({Z}_{i0}(t_{f}))]^{\top},\Gamma_{z^{k}_{i0}}=[\mathrm{vecv}(z^{k}_{i0}(t_{0}+1)),\ldots,\mathrm{vecv}(z^{k}_{i0}(t_{f}+1))]^{\top}.$
From \eqref{62}, we have
\begin{equation}
\begin{aligned}\label{65}
\tilde{\Xi}_{i}^{k}\mathrm{vecs}(\tilde{{H}}_{i}^{k})=\tilde{\varphi}_{i}^{k},
\end{aligned}
\end{equation}
where $\tilde{\varphi}_{i}^{k}=\Gamma_{{Z}_{i0}}\mathrm{vecs}(Q_{Ri})$ and $\tilde{\Xi}_{i}^{k}=\Gamma_{{Z}_{i0}}-(\gamma_{i}^{k})^{2}\Gamma_{z^{k}_{i0}}$.
Similar to Lemma \ref{L5}, if the collected data satisfy
\begin{equation}\label{66}
\begin{aligned}
\mathrm{rank}(\Gamma_{{Z}_{i0}})=(n_{i}+m_{i})(n_{i}+m_{i}+1)/2,
\end{aligned}
\end{equation}
the matrix $\tilde{\Xi}_{i}^{k}$ has full column-rank. Then, for $k=0,1,\ldots$, \eqref{65} can be solved by
\begin{equation}\label{67}
\begin{aligned}
\mathrm{vecs}(\tilde{{H}}_{i}^{k})=(\tilde{\Xi}_{i}^{k\top}\tilde{\Xi}_{i}^{k})^{-1}\tilde{\Xi}_{i}^{k\top}\tilde{\varphi}_{i}^{k}.
\end{aligned}
\end{equation}
\subsection{Determine $\tilde{\beta}_{i}$ and $\alpha_{i}^{k+1}$ by $Q$-learning.}
From Lemma \ref{L6}, if the solution $\tilde{{H}}_{i}^{k}$ to \eqref{62} is positive definite, then the system $(\tilde{\beta}_{i}+\sum_{m=0}^{k}\alpha_{i}^{m})(A_{i}-B_{i}\tilde{K}_{i}^{k})$ is stable. Then, the coefficient $\tilde{\beta}_{i}+\alpha_{i}^{0}$ can be determined by the following iterations.

{\bf Initialize} a sufficiently small constant $\alpha_{i}^{0}>0$ and a monotonic sequence $\{\tilde{\beta}_{i}^z\}$ satisfying $0<\tilde{\beta}_{i}^{z+1}<\tilde{\beta}_{i}^{z}<1$ and $\lim_{z\rightarrow\infty}\tilde{\beta}_{i}^{z}=0$ for $\forall z\in\mathbb{N}$. Set $z\leftarrow0$ and $\tilde{\beta}_{i}\leftarrow\tilde{\beta}_{i}^0$, {\bf repeat:}  1). {\bf if} $\tilde{{H}}_{i}^{0}\leq0$ from \eqref{67}, {\bf then} $z\leftarrow z+1$, $\tilde{\beta}_{i}\leftarrow\tilde{\beta}_{i}^z$ and calculate $\tilde{{H}}_{i}^{0}$ by \eqref{67}; {\bf else end}; 2). {\bf output} $\tilde{\beta}_{i}\leftarrow\tilde{\beta}_{i}^z$ and $\alpha_{i}^{0}$.

{\bf Scheme 1.} We can choose $\alpha_{i}^{k+1}$ by \eqref{44}. (Stability can be guaranteed by Lemma \ref{L7} and Theorem \ref{the3}.)

{\bf Scheme 2.} We update $\alpha_{i}^{k+1}$ by
\begin{equation}\label{76}
\begin{aligned}
\alpha_{i}^{k+1}=a_{i}\gamma^{k}_{i}
(\sqrt{\frac{\sigma_{\min}(Q_{Ri})}{\sigma_{\max}(\tilde{{H}}_{i}^{k}-Q_{Ri})}+1}-1), \quad 0<a_{i}\leq1.
\end{aligned}
\end{equation}

\begin{theorem}\label{the5}
For $i=1,\ldots,N$, given the control gain $\tilde{K}_{i}^{k+1}$ by \eqref{63}. If update $\alpha_{i}^{k+1}$ by Scheme 2, the system $\tilde{A}_{i}^{k+1}$ is Schur.
\end{theorem}

{\it Proof.}
From Lemma \ref{L6}, since $\tilde{K}_{i}^{k+1}$ is the stabilizing control gain of $\gamma_{i}^{k}(A_{i}-B_{i}\tilde{K}_{i}^{k+1})$, there is a unique positive solution $\hat{\tilde{{H}}}_{i}^{k+1}$ to $\hat{\tilde{{H}}}_{i}^{k+1} =Q_{Ri}+(\Pi_{\tilde{K}_{i}^{k+1}}^{k})^{\top}\hat{\tilde{{H}}}_{i}^{k+1}\Pi_{\tilde{K}_{i}^{k+1}}^{k}$, where $\Pi_{\tilde{K}_{i}^{k+1}}^{k}=\gamma^{k}_{i}[I_{n_{i}},(-\tilde{K}_{i}^{k+1})^{\top}]^{\top}[A_{i}, B_{i}]$. By viewing $\gamma_{i}^{k}[A_{i}, B_{i}]$ as $[A_{i},B_{i}]$ in Lemma \ref{L6a}, we have $0<\hat{\tilde{{H}}}_{i}^{k+1}\leq\tilde{{H}}_{i}^{k}$. Then, there is
\begin{equation}\label{78}
\begin{aligned}
&(\Pi_{i}^{k+1})^{\top}\hat{\tilde{{H}}}_{i}^{k+1}\Pi_{i}^{k+1}-\hat{\tilde{{H}}}_{i}^{k+1}\\
\overset{(B1)}{=}&((\gamma_{i}^{k+1}/\gamma_{i}^{k})^{2}-1)(\hat{\tilde{{H}}}_{i}^{k+1}-Q_{Ri})-Q_{Ri}\\
\overset{(C1)}{\leq}&((\gamma_{i}^{k+1}/\gamma_{i}^{k})^{2}-1)(\tilde{{H}}_{i}^{k}-Q_{Ri})-Q_{Ri}\\
\overset{(C2)}{<}&\sigma_{\min}(Q_{Ri})I_{n_{i}}-Q_{Ri}<0,
\end{aligned}
\end{equation}
where $(C1)$ is obtained by $0<\hat{\tilde{H}}_{i}^{k+1}\leq\tilde{H}_{i}^{k}$ and $(C2)$ is obtained by \eqref{76}. Similarly, any choice of $\alpha_{i}^{k+1}$ in interval \eqref{76} ensures the stability of system $\tilde{A}^{k+1}_{i}$ for step $k+1$.
$\Box$
\begin{remark}
Unlike existing PI-based $Q$-learning methods \cite{kiumarsi2014reinforcement,kiumarsi2015optimal,al2007model}, we simplify and build a new off-policy $Q$-learning framework (Lemma \ref{L6a}) by considering the case $l=0$. Moreover, the equivalence of this $Q$-learning with traditional PI is analyzed. The SPI-based $Q$-learning (Lemma \ref{L7}) is obtained by extending the SPI method to $Q$-learning, which relaxes the restriction of the initial stabilizing policy to the existing PI-based $Q$-learning.
\end{remark}
\subsection{SPI-Based $Q$-Learning COOT Algorithm 2}
\begin{algorithm}[!h]
	\caption{SPI-Based $Q$-Learning Algorithm}\label{alg2}
	\begin{algorithmic}[1]
	\STATE {\bf Initialize:} Select arbitrary control policy $\tilde{K}_{i}^{0}$, sufficiently small positive constants $\varepsilon_{i,1}$, $\varepsilon_{i,2}$ and $\alpha_{i}^{0}$, a monotonically decreasing sequence $\{\tilde{\beta}_{i}^{z}\}$, where $0<\tilde{\beta}_{i}^{z}<1$ for $\forall z$.
$k\leftarrow0$, $z\leftarrow0$, $l\leftarrow0$, $j\leftarrow0$, $\tilde{\beta}_{i}\leftarrow\tilde{\beta}_{i}^{0}$.
\STATE{\bf Pre-collection:} Set the appropriate $\mu_{i}$. Obtain $(\hat{X}_{i1},\hat{U}_{i1})$ by \eqref{16}.
\STATE {\bf Data-collection:} Collect data on the time interval $[t_{0}, t_{f}]$ until \eqref{66} and \eqref{82} are satisfied.
\STATE {\bf Online iteration:} {\it (Steps 5-13)}
\STATE  If $\tilde{{H}}_{i}^{0}\leq0$ from \eqref{67}, then $z\leftarrow z+1$, $\tilde{\beta}_{i}\leftarrow\tilde{\beta}_{i}^z$ and calculate $\tilde{{H}}_{i}^{0}\leq0$ by \eqref{67}, otherwise output $\tilde{\beta}_{i}\leftarrow\tilde{\beta}_{i}^z$ and go to next step;
\STATE Solve $\tilde{{H}}_{i}^{k}$ by \eqref{67}, update $\tilde{K}_{i}^{k+1}$ by \eqref{63};
\STATE Update $\alpha_{i}^{k+1}$ by Scheme 2 (or Scheme 1). If $\tilde{\beta}_{i}+\sum_{m=0}^{k+1}\alpha_{i}^{m}\geq1$, let $K_{i}^0\leftarrow\tilde{K}_{i}^{k+1}$ and go to the next step; Otherwise $k\leftarrow k+1$ and go to Step 6;
\STATE Solve ${{H}}_{i}^{j}$ by \eqref{79}, update control policy $K_{i}^{j+1}$ by \eqref{69};
\STATE If $\|{{H}}_{i}^{j}-{{H}}_{i}^{j-1}\|\leq\varepsilon_{i,1}$, go to the next step; Otherwise $j\leftarrow j+1$ and go to Step 8;
\STATE Solve \eqref{83} for $l=1,2,\ldots,h_{i}$ to obtain $\Omega_{i}$ and $\hat{\eta}_{i}$;
\STATE {Choose parameter $0<\kappa_{i}<2/\rho(\Omega_{i}^{\top}\Omega_{i})$;}
\STATE {Calculate $\hat{\chi}_{i}^{n}$ by \eqref{20}; }
\STATE {If $\|\hat{\chi}_{i}^{n+1}-\hat{\chi}_{i}^{n}\|\leq\varepsilon_{i,2}$, output $\hat{X}_{i}$ and $\hat{U}_{i}$ and go to the next step; Otherwise $n\leftarrow n+1$ and go to Step 12.}
\STATE {\bf Optimal control phase:}  Set $K_{i}\leftarrow K_{i}^{j}$ and ${T}_{i}\leftarrow \hat{U}_{i}+K^{j}_{i}\hat{X}_{i}$ and use the optimal controller \eqref{4}.
	\end{algorithmic}
\end{algorithm}
By using \eqref{65}, \eqref{67} and Scheme 2 (or Scheme 1), we obtain the stabilizing control gain $\tilde{K}_{i}^{k+1}$ if $\tilde{\beta}_{i}+\sum_{m=0}^{k+1}\alpha_{i}^{m}\geq1$ is satisfied. Let $K_{i}^0=\tilde{K}_{i}^{k+1}$ be the initial stabilizing gain for $i=1,\ldots,N$. If \eqref{66} is satisfied, for $j=0,1,\ldots$, \eqref{68} can be solved by
\begin{equation}\label{79}
\begin{aligned}
\mathrm{vecs}({H}_{i}^{j})=({\Xi}_{i}^{j\top}{\Xi}_{i}^{j})^{-1}{\Xi}_{i}^{j\top}{\varphi}_{i}^{j},
\end{aligned}
\end{equation}
where ${\Xi}_{i}^{j}=\Gamma_{{Z}_{i0}}-\Gamma_{z^{j}_{i0}}$
and ${\varphi}_{i}^{j}=\Gamma_{{Z}_{i0}}\mathrm{vecs}(Q_{Ri})$ with $z^{j}_{i0}(t+1)=[\tilde{x}_{i0}(t+1)^{\top},({K}_{i}^{j}\tilde{x}_{i0}(t+1))^{\top}]^{\top}$.

Based on ${H}_{i}^{j}$ and ${K}_{i}^{j+1}$ obtained from \eqref{79} and \eqref{69}, we further solve \eqref{19b} for $l=1,\ldots,h_{i}$. According to Lemma \ref{L6a} and  ${P}_{i}^{j}=[I_{n_{i}},(-{K}_{i}^{j})^{\top}]{{H}}^{j}_{i}[I_{n_{i}},(-{K}_{i}^{j})^{\top}]^{\top}$, one has that
$[\tilde{x}_{il}^{\top}(t+1),
-({K}^{j}_{i}\tilde{x}_{il}(t+1))^{\top}]{H}_{i}^{j}[
\tilde{x}_{il}^{\top}(t+1),
-({K}^{j}_{i}\tilde{x}_{il}(t+1))^{\top}
]^{\top}
-
[\tilde{x}_{il}^{\top}(t),
-({K}^{j}_{i}\tilde{x}_{il}(t))^{\top}]{H}_{i}^{j}[
\tilde{x}_{il}^{\top}(t),
-({K}^{j}_{i}\tilde{x}_{il}(t))^{\top}
]^{\top}
=\tilde{x}_{il}^{T}(t+1){P}_{i}^{j}\tilde{x}_{il}(t+1)-\tilde{x}_{il}^{T}(t){P}_{i}^{j}\tilde{x}_{il}(t)$.
And \eqref{29} can be rewritten as
$\tilde{x}_{il}(t+1)={A}_{i}^{j}\tilde{x}_{il}(t)
+B_{i}({K}_{i}^{j}\tilde{x}_{il}(t)+u_{i}(t))+\pi_{il}\zeta_{i}(t)
+\varrho_{il}(t)$,
where ${A}_{i}^{j}=A_{i}-B_{i}K_{i}^{j}$. Since $\|\varrho_{il}(t_{0})\|<\vartheta_{i,4}$, if $\vartheta_{i,4}$ is sufficiently small, by substituting the above equation to $\tilde{x}_{il}^{\top}(t+1){P}_{i}^{j}\tilde{x}_{il}(t+1)-\tilde{x}_{il}^{\top}(t){P}_{i}^{j}\tilde{x}_{il}(t)$ and using the data matrices in Section \ref{sec4.1.1}, we have
\begin{equation}\label{81}
\begin{aligned}
\Phi_{il}[\mathrm{vec}^{\top}({L}_{3,il}^{j}),\mathrm{vec}^{\top}({L}_{4,il}^{j}),\mathrm{vecs}^{\top}({L}_{5,il}^{j})]^{\top}=\hbar_{il},
\end{aligned}
\end{equation}
where $\hbar_{il}=\theta_{\tilde{x}_{il}}\mathrm{vecs}(P_{i}^{j})-(2\Gamma_{\tilde{x}_{il}}(I_{n_{i}}\otimes{K}_{i}^{j\top})+2\Gamma_{u_{i}\tilde{x}_{il}})\mathrm{vec}(H_{i,12}^j)
+(\Gamma_{\tilde{x}\tilde{x}_{il}}-\Gamma_{u_{i}})\mathrm{vecs}(H_{i,22}^{j}-R_{i})+\Gamma_{\tilde{x}\tilde{x}_{il}}\mathrm{vecs}(Q_{i}+{K}_{i}^{j\top}R_{i}{K}_{i}^{j})$ and $\Phi_{il}=[2{\Gamma}_{\zeta_{i}\tilde{x}_{il}},2\Gamma_{\zeta_{i}u_{i}},\Gamma_{\zeta_{i}}]$.

If the condition
\begin{equation}\label{82}
\begin{aligned}
\mathrm{rank}([{\Gamma}_{\zeta_{i}\tilde{x}_{il}},\Gamma_{\zeta_{i}{u}_{i}},\Gamma_{\zeta_{i}}])=(n_{i}+m_{i})n_{v}+n_{v}(n_{v}+1)/2
\end{aligned}
\end{equation}
is satisfied for $l=1,2,\ldots,h_{i}$, then \eqref{81} can be solved by
\begin{equation}\label{83}
\begin{aligned}
&[\mathrm{vec}^{\top}({L}_{3,il}^{j}),\mathrm{vec}^{\top}({L}_{4,il}^{j}),\mathrm{vecs}^{\top}({L}_{5,il}^{j})]^{\top}
=(\Phi_{il}^{\top}\Phi_{il})^{-1}\Phi_{il}^{\top}\hbar_{il}.
\end{aligned}
\end{equation}
Based on \eqref{83}, let $M_{i}=A_{i}^{\top}P_{i}^{j}$ in \eqref{19b}. We have $M_{i}\bar{\Upsilon}_{i}(X_{il},U_{il})=-A_{i}^{\top}P_{i}^{j}\pi_{il}-A_{i}^{\top}P_{i}^{j}B_{i}U_{il}
=-{L}_{3,il}^{j}-H_{i,12}^{j}U_{il}$. The $\Omega_{i}$ and $\hat{\eta}_{i}$ in \eqref{19b} can be obtained by using
$M_{i}\bar{\Upsilon}_{i}(X_{il},U_{il})=-{L}_{3,il}^{j}-H_{i,12}^{j}U_{il}$.

\begin{theorem}\label{the6}
Under Assumptions \ref{A1}-\ref{A4}, consider the distributed observer \eqref{3} and system \eqref{1a}-\eqref{2}. For $i=1,\ldots,N$, use the adaptive feedforward-feedback controller \eqref{4} with ${K}_{i}^{j}$ and ${T}_{i}$ being the optimal gains learned by Algorithm \ref{alg2}. Then, the COOT problem is solved.
\end{theorem}

{\it Proof.}
For $i=1,\ldots,N$, if \eqref{66} is satisfied, \eqref{65} can be solved by \eqref{67}. By Lemma \ref{L7} and Theorem \ref{the3}, choosing $\alpha_{i}^{k+1}$ by Scheme 1 satisfies \eqref{28}. By Theorem \ref{the5}, choosing $\alpha_{i}^{k+1}$ by Scheme 2 can ensure the stability of system $\tilde{A}_{i}^{k+1}$ and satisfy \eqref{28}. Since \eqref{65} is equivalent to \eqref{62}, Steps 5-7 in Algorithm \ref{alg2} satisfy Lemma \ref{L7}. If $\tilde{\beta}_{i}+\sum_{m=0}^{k+1}\alpha_{i}^{m}\geq1$ is satisfied, then $\tilde{K}_{i}^{k+1}$ is the stabilizing control gain of $A_{i}-B_{i}\tilde{K}_{i}^{k+1}$. Let $K_{i}^0=\tilde{K}_{i}^{k+1}$ be the stabilizing policy, which satisfies the initial condition of Lemma \ref{L6a}. For $j=0,1,\ldots$, if \eqref{66} is satisfied, \eqref{68} can be solved by \eqref{79}. Then, Steps 8-9 in Algorithm \ref{alg2} are equivalent to Lemma \ref{L6a}. Therefore, we can obtain $\lim_{j\rightarrow\infty}{H}_{i}^{j}={H}_{i}^{*}$ and $\lim_{j\rightarrow\infty}K_{i}^{j}=K_{i}^{*}$ by Algorithm \ref{alg2}. If $\varrho_{il}(t_{0})\rightarrow0$ and \eqref{82} is satisfied, \eqref{81} can be solved by \eqref{83}. The $\Omega_{i}$ and $\hat{\eta}_{i}$ in \eqref{19b} can be obtained. By Theorem \ref{the2}, $\hat{\chi}_{i}^{n}$ can converge to the solution of \eqref{19a} and thus $(X_{i},U_{i})$ is obtained. From Theorem \ref{the1} and $\lim_{j\rightarrow\infty}K_{i}^{j}=K_{i}^{*}$, we have $\lim_{t\rightarrow\infty}e_{i}(t)=0$. Thus, the COOT problem is solved.$\Box$
\begin{remark}
By separating the process of solving \eqref{79} and \eqref{81}, we obtain two different data conditions \eqref{66} and \eqref{82}. 
In Algorithm \ref{alg1}, the full column-rank condition is \eqref{37}. Define $\mathcal{R}_{i,1}=\mathrm{rank}([\Gamma_{\tilde{x}\tilde{x}_{il}},\Gamma_{u_{i}},\Gamma_{\zeta_{i}},\Gamma_{u_{i}\tilde{x}_{il}},\Gamma_{\zeta_{i}\tilde{x}_{il}},\Gamma_{\zeta_{i}u_{i}}])
=(n_{i}+m_{i}+n_{v})(n_{i}+m_{i}+n_{v}+1)/2$. Define ${r}_{i,1}=\mathrm{rank}(\Gamma_{{Z}_{i0}})=(n_{i}+m_{i})(n_{i}+m_{i}+1)/2$ and ${r}_{i,2}=\mathrm{rank}([{\Gamma}_{\zeta_{i}\tilde{x}_{il}},\Gamma_{\zeta_{i}{u}_{i}},\Gamma_{\zeta_{i}}])=(n_{i}+m_{i})n_{v}+n_{v}(n_{v}+1)/2$.
Then, the rank that satisfies Algorithm \ref{alg2} is $\mathcal{R}_{i,2}=\max\{{r}_{i,1},{r}_{i,2}\}$. Since $\mathcal{R}_{i,1}={r}_{i,1}+{r}_{i,2}$, it follows that $\mathcal{R}_{i,1}>\mathcal{R}_{i,2}$. If $t_{f1}$ is defined as the time when condition \eqref{37} is satisfied and $t_{f2}$ as the time when \eqref{66} and \eqref{82} are satisfied, then there is $t_{f1}>t_{f2}$. Clearly, Algorithm \ref{alg2} has a lower data full column-rank condition.
\end{remark}
\begin{remark}
Different from Algorithm \ref{alg2}, Algorithm \ref{alg1} is an online iteration algorithm based on Lyapunov functions, which updates the control policy in each step according to the obtained stabilizing Lyapunov solution, and is more robust and can be extended to more complex problems, such as the cooperative optimal output regulation \cite{xie2023data}. Algorithm \ref{alg1} solves the initial stabilizing policy limitations in Lyapunov-based off-policy methods \cite{chen2022robust,kiumarsi2015optimal,jiang2019optimal}. Moreover, even if VI \cite{li2022model,kiumarsi2015optimal,xie2023data} or hybrid iteration \cite{10167108} are used to realize COOT without an initial stabilizing policy, they data conditions required are not less than the condition in Algorithm \ref{alg1}.
\end{remark}

\section{Simulation}\label{section:6}
\begin{figure*}[htbp]
    \centering
	  \subfloat[]{
       \includegraphics[width=0.2\linewidth]{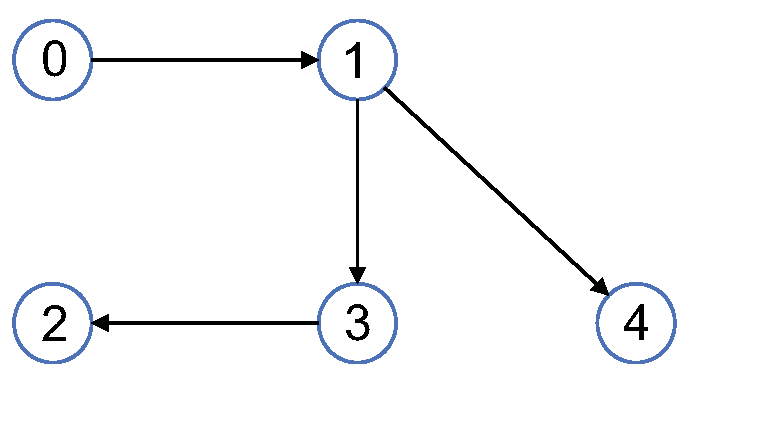}}
       \hspace{-10pt}
	  \subfloat[]{
        \includegraphics[width=0.2\linewidth]{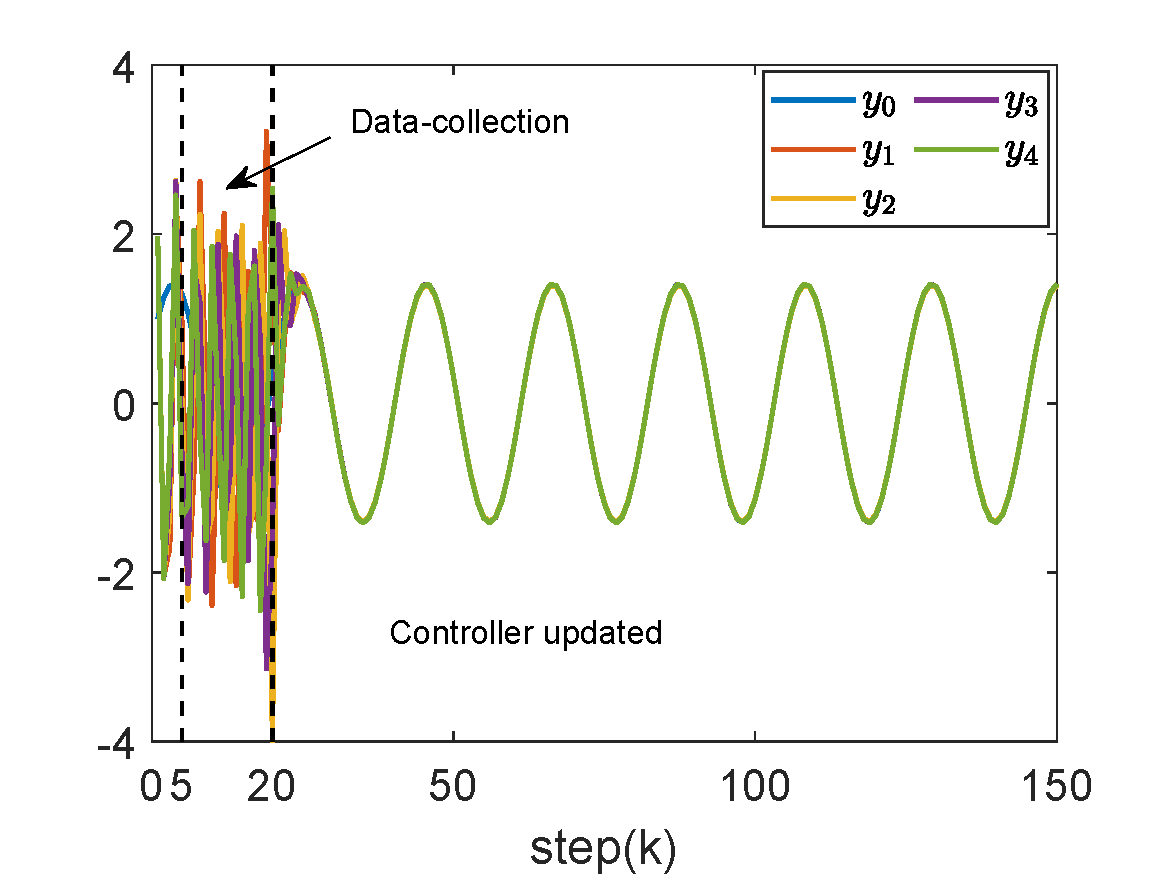}}
         \hspace{-10pt}
	  \subfloat[]{
        \includegraphics[width=0.2\linewidth]{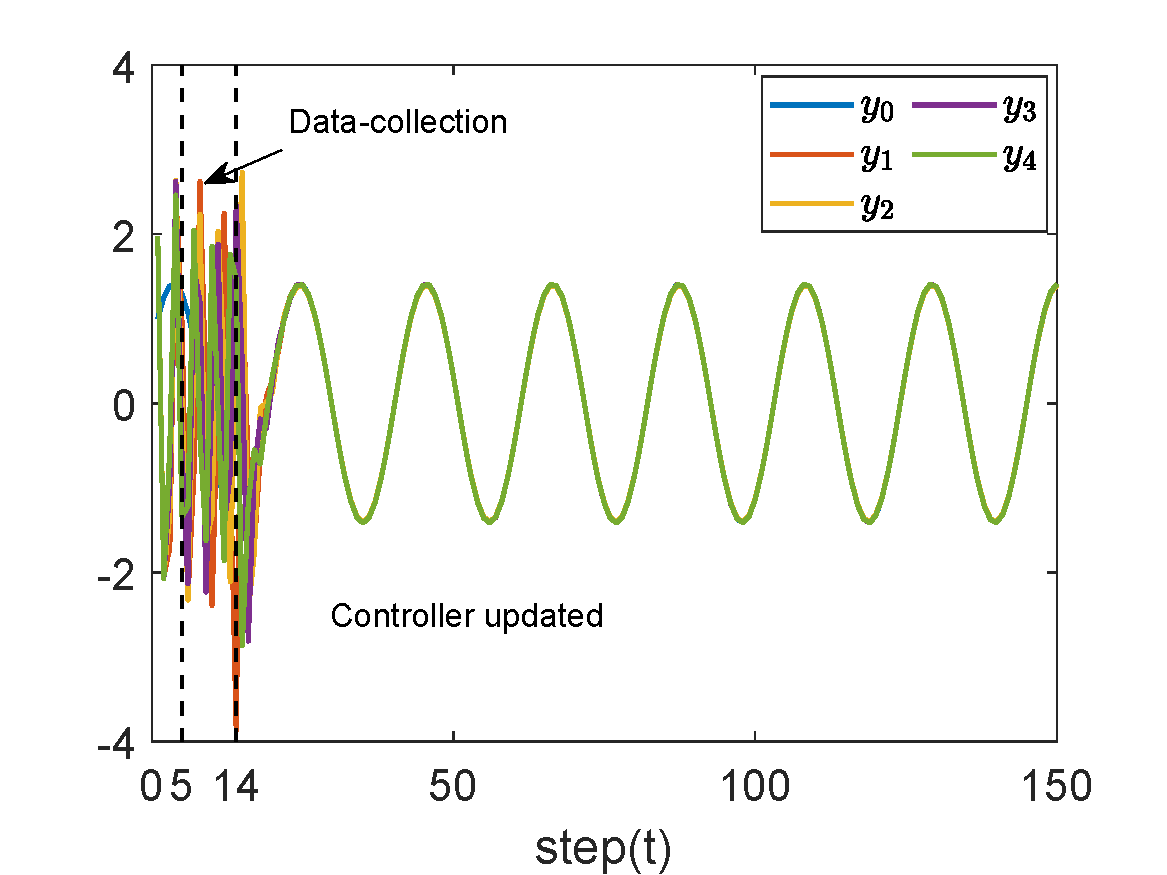}}
         \hspace{-10pt}
	  \subfloat[]{
        \includegraphics[width=0.2\linewidth]{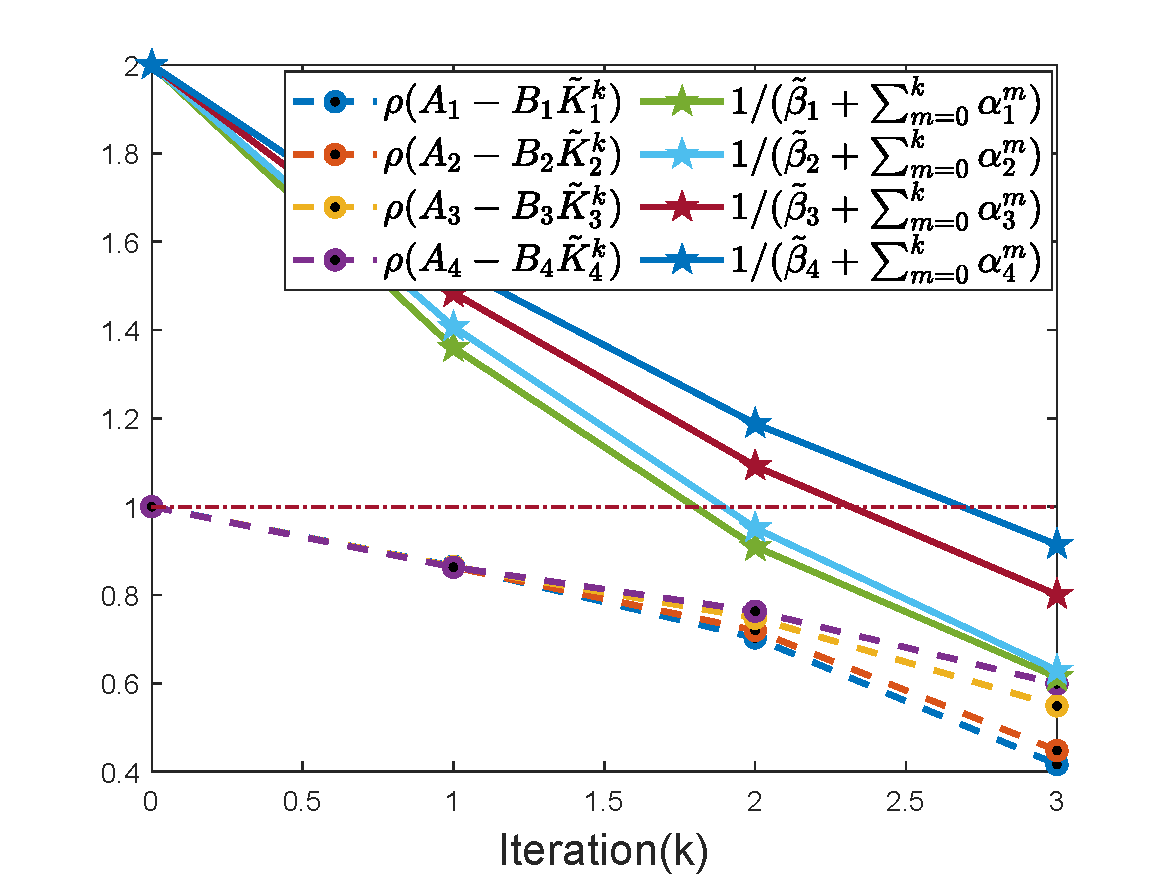}}
         \hspace{-10pt}
	  \subfloat[]{
        \includegraphics[width=0.2\linewidth]{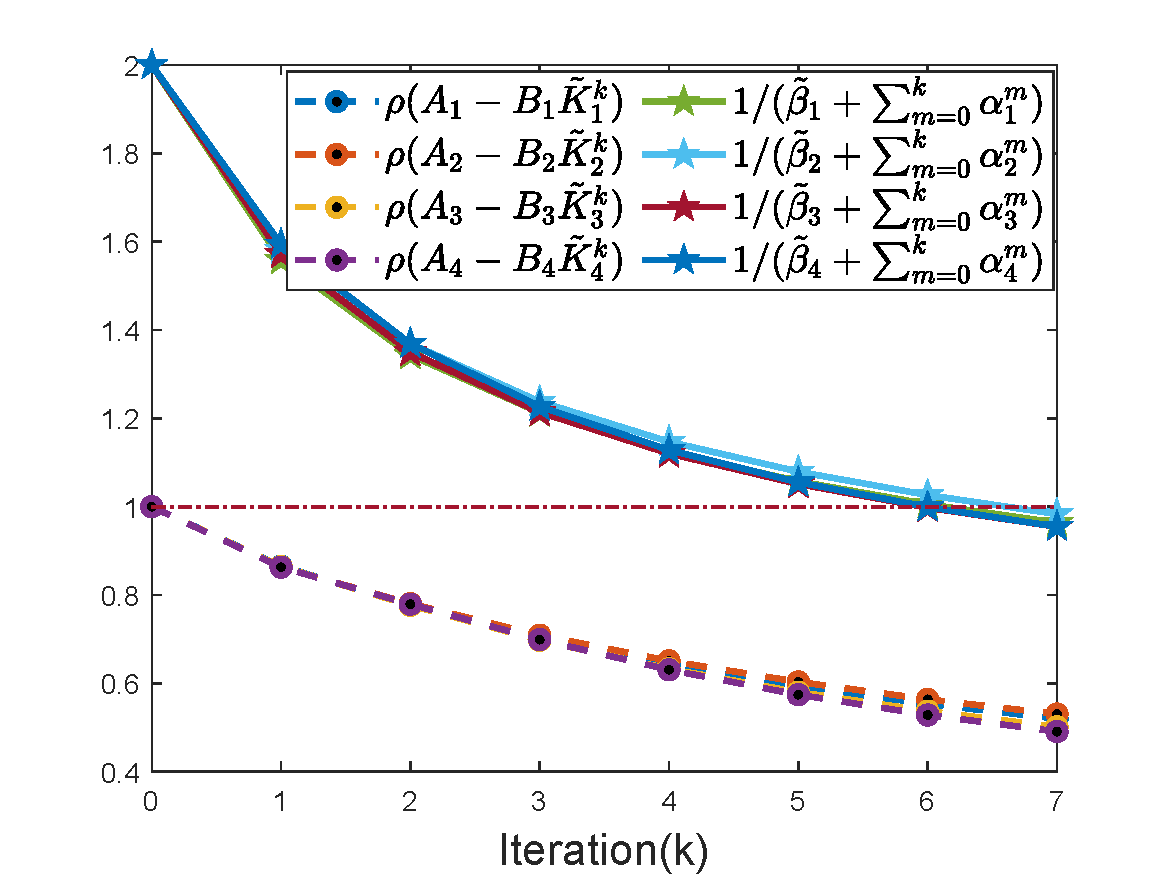}}
         \hspace{-10pt}
         \subfloat[]{
        \includegraphics[width=0.2\linewidth]{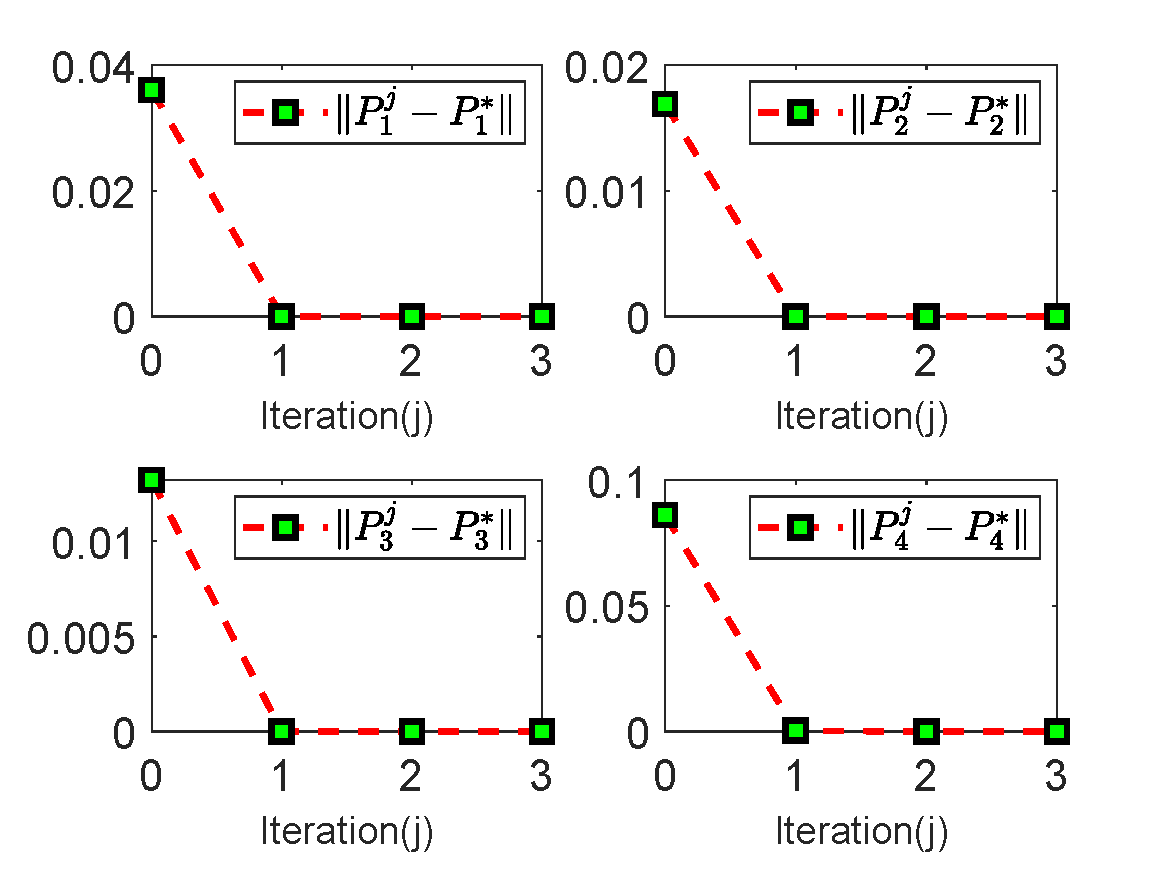}}
         \hspace{-10pt}
         \subfloat[]{
        \includegraphics[width=0.2\linewidth]{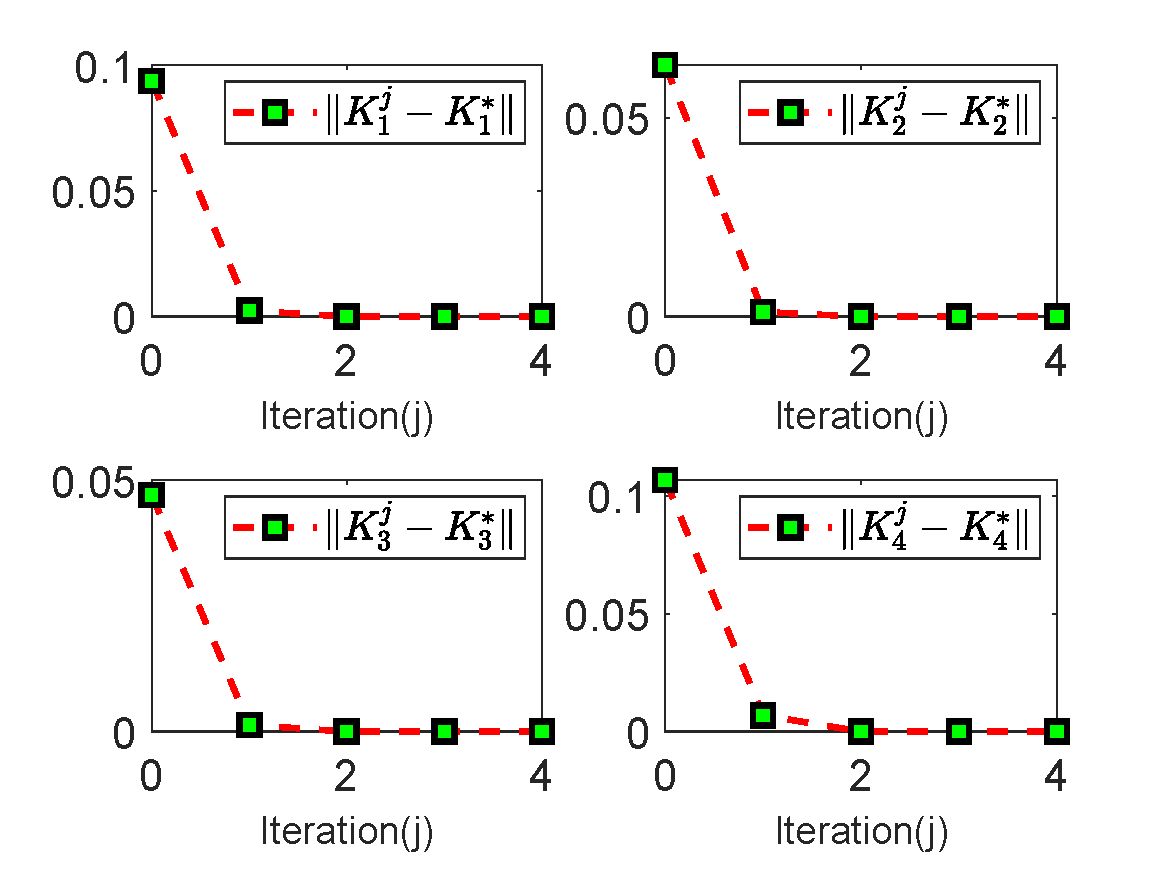}}
         \hspace{-10pt}
         \subfloat[]{
        \includegraphics[width=0.2\linewidth]{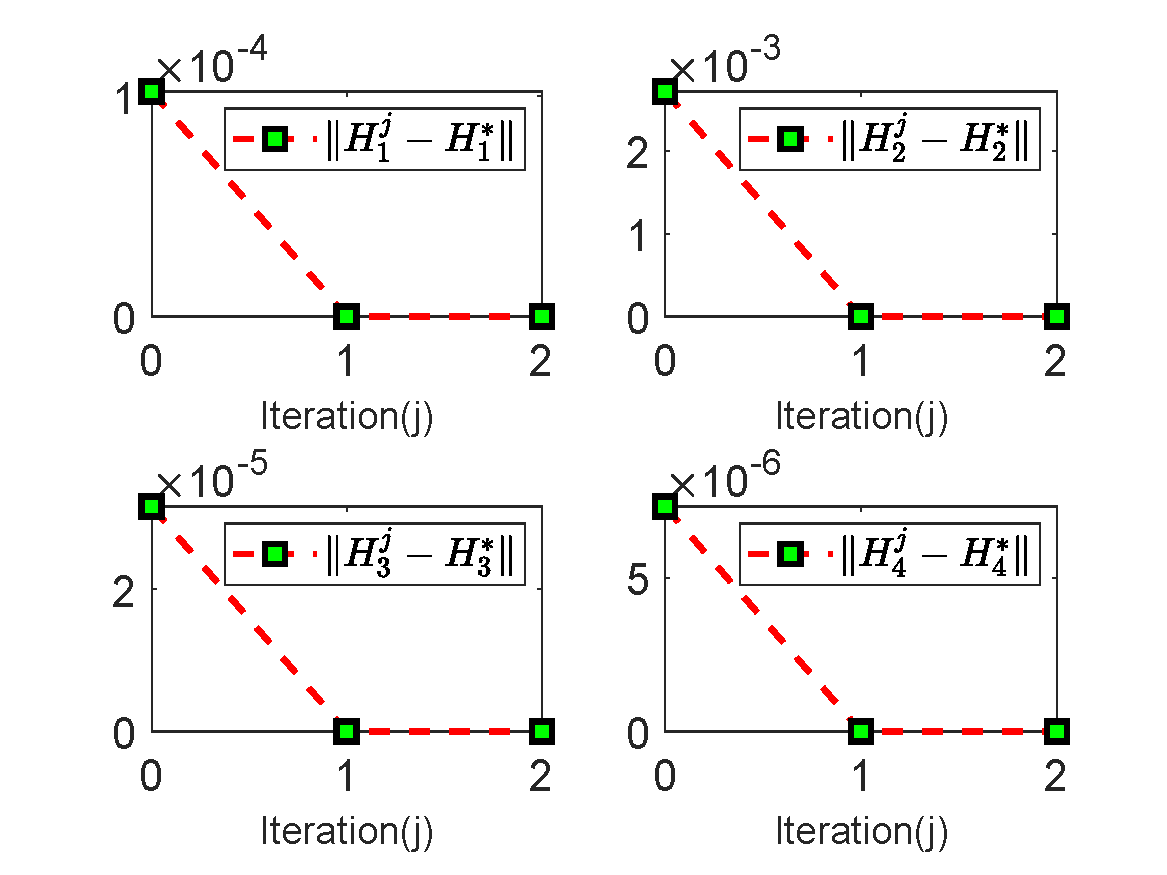}}
        \hspace{-10pt}
         \subfloat[]{
        \includegraphics[width=0.2\linewidth]{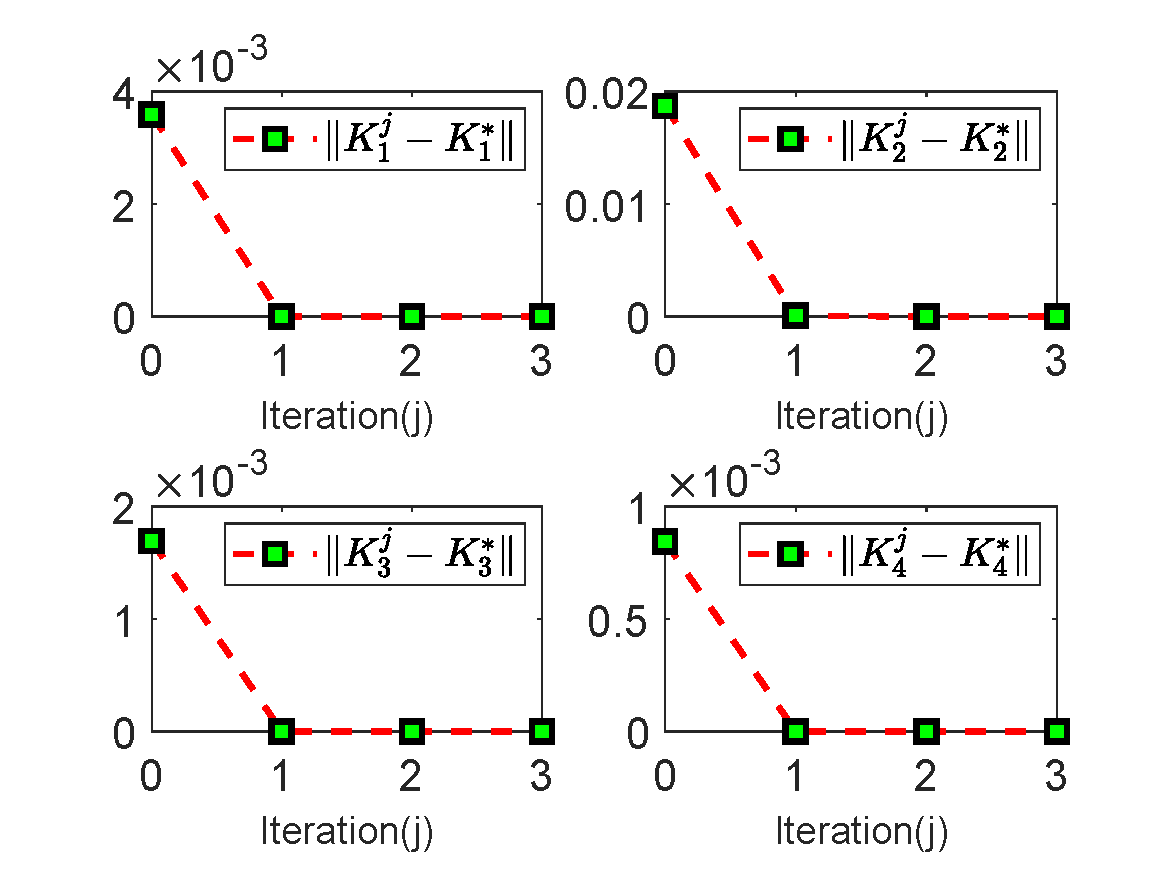}}
        \hspace{-10pt}
         \subfloat[]{
        \includegraphics[width=0.2\linewidth]{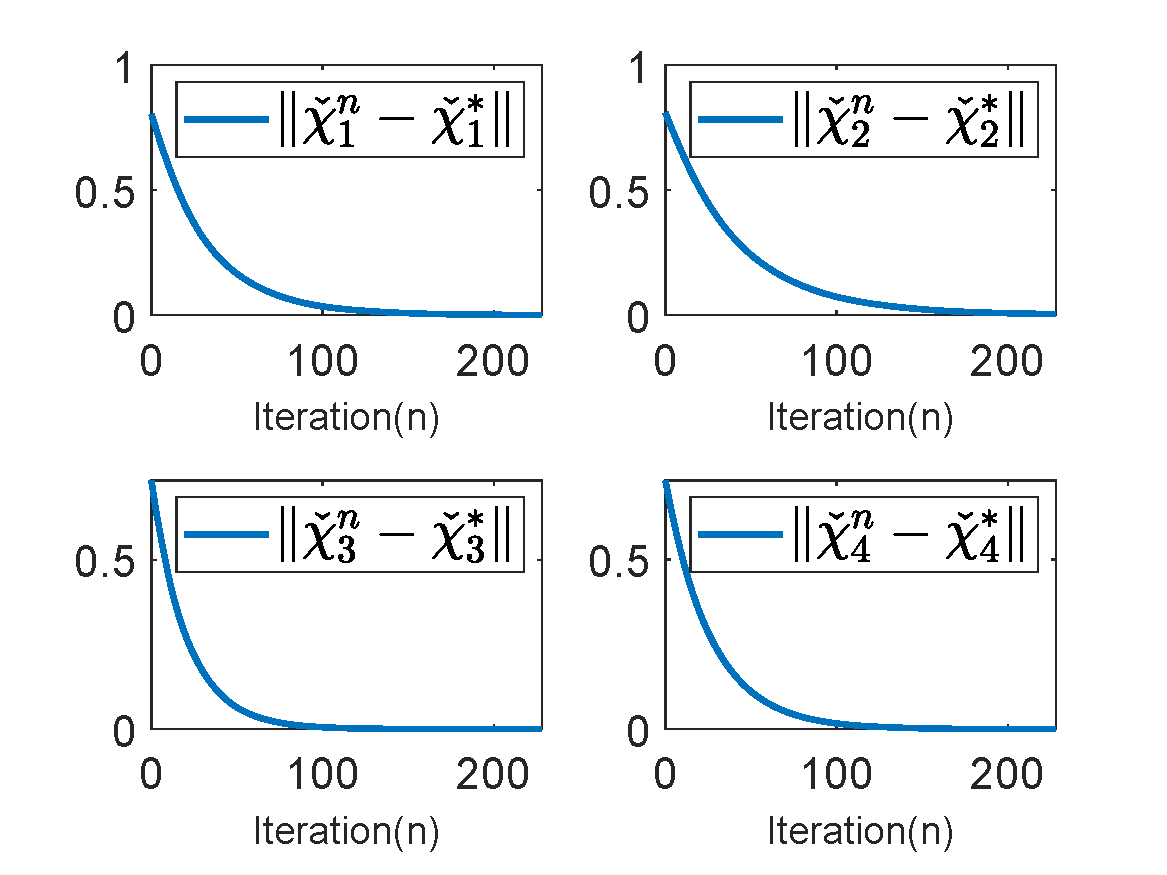}}
	  \caption{(a). The communication topology; (b). Output tracking results of Algorithm \ref{alg1}; (c). Output tracking results of Algorithm \ref{alg2}; (d). The closed-loop spectral radius by Algorithm \ref{alg1} with Scheme 1; (e). The closed-loop spectral radius by Algorithm \ref{alg2} with Scheme 2; (f). $\|P_{i}^{j}-P_{i}^*\|$ of Algorithm \ref{alg1}; (g). $\|H_{i}^{j}-H_{i}^*\|$ of Algorithm \ref{alg2}; (h). $\|K_{i}^{j}-K_{i}^*\|$ of Algorithm \ref{alg1}; (i). $\|K_{i}^{j}-K_{i}^*\|$ of Algorithm \ref{alg2}; (j). The iterative solutions to \eqref{6} by Algorithms \ref{alg1} and \ref{alg2}.}\label{fig1}
\end{figure*}
In this section, Algorithms \ref{alg1} and \ref{alg2} will be verified.
Set
$E=\left[
\begin{matrix}
\cos(0.3)&\sin(0.3)\\
-\sin(0.3)&\cos(0.3)
\end{matrix}
\right]$,
$A_{i}=\left[
\begin{matrix}
0&1\\
-1&-0.2i
\end{matrix}
\right]$, $F=[-1,0]$, $C_{i}=[1,0]$ and $S_{i}=1$ for $i=1,2,3,4$, $B_{1,2}=[0,1]^{\top}$ and $B_{3,4}=[1,0]^{\top}$. Since $\rho(A_{i})\geq1$, all 4 followers are open-loop unstable. The MASs are 4 followers and 1 leader connected by the communication graph in Fig. \ref{fig1}(a).
Set $Q_{i}=I_{2}$ and $R_{i}=1$. The initial states are set to $v(0)=[1,1]^{\top}$, $x_{i}(0)=[2,-2]^{\top}$, $\zeta_{i}(0)=[0,0]^{\top}$, $E_{i}(0)={\bf 0}_{2\times2}$ and $F_{i}(0)=[0,0]$. When $A_{i}$ and $B_{i}$ are unknown, set $\alpha_{i}^0=10^{-4}$ and the monotonically decreasing sequence $\tilde{\beta}^{z}_{i}$ as $\{0.5,0.49,0.48,\ldots,0.01\}$.  For Algorithms \ref{alg1} and \ref{alg2}, we set the exploration noises as $n_{i}(t)=0.1\sin(16t)+0.1\cos(11t)$, the initial time for collecting data as $t_{0}=5$, and select $\mu_{i}=1/\rho(\mathcal{H})=1$, $\kappa_{i}=1/\rho(\Omega_{i}^{\top}\Omega_{i})$, $a_{i}=0.5$ and $\varepsilon_{i,1}=\varepsilon_{i,2}=10^{-4}$.

{\it For Algorithm \ref{alg1}.} There is $t_{f}=20$ step. The output tracking results are shown in Fig. \ref{fig1} (b). The closed-loop spectral radius during the calculation of the stabilizing control gain is illustrated in Fig. \ref{fig1} (d). The optimality error $\|P_{i}^{j}-P_{i}^*\|$ is demonstrated in Fig. \ref{fig1} (f), and the optimal control error $\|K_{i}^{j}-K_{i}^*\|$ is shown in Fig. \ref{fig1} (g).

{\it For Algorithm \ref{alg2}.} The output tracking results are shown in Fig. \ref{fig1} (c), where $t_{f}=14$. This validates the lower data condition advantage of Algorithm \ref{alg2}. The closed-loop spectral radius is illustrated in Fig. \ref{fig1} (e). The optimality error $\|H_{i}^{j}-H_{i}^*\|$ is demonstrated in Fig. \ref{fig1} (h), and $\|K_{i}^{j}-K_{i}^*\|$ is shown in Fig. \ref{fig1} (i).

We obtain the same result as shown in Fig. \ref{fig1} (j), where $\check{\chi}_{i}^{n}=\mathrm{vec}([(\hat{X}_{i}^{n})^{\top},(\hat{U}_{i}^{n})^{\top}]^{\top})$ and $\check{\chi}_{i}^{*}=\mathrm{vec}([(X_{i}^{*})^{\top},(U_{i}^{*})^{\top}]^{\top})$. The effectiveness of Algorithms \ref{alg1} and \ref{alg2} is verified by the above results.

\section{Conclusion}\label{section:7}
In this paper, two data-driven stabilizing COOT algorithms are proposed. First, a model-based SPI framework is proposed for finding stabilizing control policies. Then, the model-based approach is extended to two data-driven online learning versions, i.e., Lyapunov-based off-policy SPI and SPI-based $Q$-learning. Finally, the proposed algorithms are validated by simulation.


\begin{thebibliography}{10}
\providecommand{\url}[1]{#1}
\csname url@samestyle\endcsname
\providecommand{\newblock}{\relax}
\providecommand{\bibinfo}[2]{#2}
\providecommand{\BIBentrySTDinterwordspacing}{\spaceskip=0pt\relax}
\providecommand{\BIBentryALTinterwordstretchfactor}{4}
\providecommand{\BIBentryALTinterwordspacing}{\spaceskip=\fontdimen2\font plus
\BIBentryALTinterwordstretchfactor\fontdimen3\font minus
  \fontdimen4\font\relax}
\providecommand{\BIBforeignlanguage}[2]{{%
\expandafter\ifx\csname l@#1\endcsname\relax
\typeout{** WARNING: IEEEtran.bst: No hyphenation pattern has been}%
\typeout{** loaded for the language `#1'. Using the pattern for}%
\typeout{** the default language instead.}%
\else
\language=\csname l@#1\endcsname
\fi
#2}}
\providecommand{\BIBdecl}{\relax}
\BIBdecl

\bibitem{lewis2010reinforcement}
F.~L. Lewis and K.~G. Vamvoudakis, ``Reinforcement learning for partially
  observable dynamic processes: Adaptive dynamic programming using measured
  output data,'' \emph{IEEE Trans. Syst. Man Cybern. Part B-Cybern.}, vol.~41,
  no.~1, pp. 14--25, 2010.

\bibitem{lewis2012optimal}
F.~L. Lewis, D.~Vrabie, and V.~L. Syrmos, \emph{Optimal control}.  John Wiley \& Sons, 2012.


\bibitem{luo2020policy}
B.~Luo, Y.~Yang, and D.~Liu, ``Policy iteration \emph{{Q}}-learning for
  data-based two-player zero-sum game of linear discrete-time systems,''
  \emph{IEEE T. Cybern.}, vol.~51, no.~7, pp. 3630--3640, 2020.

\bibitem{li2024data}
D.~Li and J.~Dong, ``Data-based efficient off-policy stabilizing optimal
  control algorithms for discrete-time linear systems via damping
  coefficients,'' \emph{arXiv preprint arXiv:2412.20845}, 2024.

\bibitem{hewer1971iterative}
G.~Hewer, ``An iterative technique for the computation of the steady state
  gains for the discrete optimal regulator,'' \emph{IEEE Trans. Autom.
  Control}, vol.~16, no.~4, pp. 382--384, 1971.

\bibitem{chen2022robust}
C.~Chen, L.~Xie, Y.~Jiang, K.~Xie, and S.~Xie, ``Robust output regulation and
  reinforcement learning-based output tracking design for unknown linear
  discrete-time systems,'' \emph{IEEE Trans. Autom. Control}, vol.~68, no.~4,
  pp. 2391--2398, 2023.

\bibitem{kiumarsi2014reinforcement}
B.~Kiumarsi, F.~L. Lewis, H.~Modares, A.~Karimpour, and M.-B. Naghibi-Sistani,
  ``Reinforcement \emph{{Q}}-learning for optimal tracking control of linear
  discrete-time systems with unknown dynamics,'' \emph{Automatica}, vol.~50,
  no.~4, pp. 1167--1175, 2014.

\bibitem{kiumarsi2015optimal}
B.~Kiumarsi, F.~L. Lewis, M.-B. Naghibi-Sistani, and A.~Karimpour, ``Optimal
  tracking control of unknown discrete-time linear systems using input-output
  measured data,'' \emph{IEEE Trans. Cybern.}, vol.~45, no.~12, pp. 2770--2779,
  2015.

\bibitem{jiang2019optimal}
Y.~Jiang, B.~Kiumarsi, J.~Fan, T.~Chai, J.~Li, and F.~L. Lewis, ``Optimal
  output regulation of linear discrete-time systems with unknown dynamics using
  reinforcement learning,'' \emph{IEEE Trans. Cybern.}, vol.~50, no.~7, pp.
  3147--3156, 2020.

\bibitem{lopez2023efficient}
V.~G. Lopez, M.~Alsalti, and M.~A. M{\"u}ller, ``Efficient off-policy
  \emph{{Q}}-learning for data-based discrete-time $\mathrm{{LQR}}$ problems,''
  \emph{IEEE Trans. Autom. Control}, vol.~68, no.~5, pp. 2922--2933, 2023.

\bibitem{al2007model}
A.~Al-Tamimi, F.~L. Lewis, and M.~Abu-Khalaf, ``Model-free \emph{{Q}}-learning
  designs for linear discrete-time zero-sum games with application to
  \emph{{H}}-infinity control,'' \emph{Automatica}, vol.~43, no.~3, pp. 473--481, 2007.

\bibitem{li2022model}
C.~Li, J.~Ding, F.~L. Lewis, and T.~Chai, ``Model-free \emph{{Q}}-learning for
  the tracking problem of linear discrete-time systems,'' \emph{IEEE Trans.
  Neural Netw. Learn. Syst.}, vol.~35, no.~3, pp. 3191--3201, 2024.

\bibitem{gao2018leader}
W.~Gao, Z.-P. Jiang, F.~L. Lewis, and Y.~Wang, ``Leader-to-formation stability
  of multiagent systems: An adaptive optimal control approach,'' \emph{IEEE
  Trans. Autom. Control}, vol.~63, no.~10, pp. 3581--3587, 2018.

\bibitem{chen2020off}
C.~Chen, F.~L. Lewis, K.~Xie, S.~Xie, and Y.~Liu, ``Off-policy learning for
  adaptive optimal output synchronization of heterogeneous multi-agent
  systems,'' \emph{Automatica}, vol. 119, p. 109081, 2020.

\bibitem{jiang2023reinforcement}
Y.~Jiang, W.~Gao, J.~Wu, T.~Chai, and F.~L. Lewis, ``Reinforcement learning and
  cooperative \emph{{H}}$_\infty$ output regulation of linear continuous-time
  multi-agent systems,'' \emph{Automatica}, vol. 148, p. 110768, 2023.

\bibitem{chen2023distributed}
C.~Chen, F.~L.~Lewis, K.~Xie, Y.~Lyu, and S.~Xie, ``Distributed output data-driven
 optimal robust synchronization of heterogeneous multi-agent
 systems," \emph{Automatica}, vol. 153, p. 111030, 2023.


\bibitem{xie2023data}
K.~Xie, Y.~Jiang, X.~Yu, and W.~Lan, ``Data-driven cooperative optimal output
  regulation for linear discrete-time multi-agent systems by online distributed
  adaptive internal model approach,'' \emph{Sci. China-Inf. Sci.}, vol.~66,
  no.~7, p. 170202, 2023.

\bibitem{huang2004nonlinear}
J.~Huang, \emph{Nonlinear output regulation: theory and applications}. SIAM, 2004.

\bibitem{luo2019balancing}
B.~Luo, Y.~Yang, H.-N. Wu, and T.~Huang, ``Balancing value iteration and policy
  iteration for discrete-time control,'' \emph{IEEE Trans. Syst. Man Cybern.
  -Syst.}, vol.~50, no.~11, pp. 3948--3958, 2020.

\bibitem{lamperski2020computing}
A.~Lamperski, ``Computing stabilizing linear controllers via policy
  iteration,'' in \emph{2020 59th IEEE Conf. Decis. Control (CDC)}, 2020, pp.
  1902--1907.

\bibitem{chen2022homotopic}
C.~Chen, F.~L. Lewis, and B.~Li, ``Homotopic policy iteration-based learning
  design for unknown linear continuous-time systems,'' \emph{Automatica}, vol.
  138, p. 110153, 2022.

\bibitem{li2024off}
D.~Li and J.~Dong, ``Off-policy learning \emph{H}$_{\infty}$ cooperative
optimal output regulation for unknown multiagent systems without initial
stabilizing gains," \emph{IEEE Trans. Ind. Inform.}, vol.20, no.11, pp.13441--13451,2024.

\bibitem{jiang2022bias}
H.~Jiang and B.~Zhou, ``Bias-policy iteration based adaptive dynamic
  programming for unknown continuous-time linear systems,'' \emph{Automatica},
  vol. 136, p. 110058, 2022.

\bibitem{gao2022resilient}
W.~Gao, C.~Deng, Y.~Jiang, and Z.-P. Jiang, ``Resilient reinforcement learning
  and robust output regulation under denial-of-service attacks,''
  \emph{Automatica}, vol. 142, p. 110366, 2022.

\bibitem{10167108}
O.~Qasem, W.~Gao, and H.~Gutierrez, ``Adaptive optimal control for
  discrete-time linear systems via hybrid iteration,'' in \emph{2023 IEEE 12th
  Data Driven Control Learn. Syst. Conf. (DDCLS)}, 2023, pp. 1141--1146.

\bibitem{huang2016cooperative}
J.~Huang, ``The cooperative output regulation problem of discrete-time linear
  multi-agent systems by the adaptive distributed observer,'' \emph{IEEE Trans.
  Autom. Control}, vol.~62, no.~4, pp. 1979--1984, 2016.


\bibitem{jiang2024adaptive}
Y.~Jiang, L.~Liu, and G.~Feng, ``Adaptive optimal tracking control of networked
  linear systems under two-channel stochastic dropouts,'' \emph{Automatica},
  vol. 165, p. 111690, 2024.

\end{thebibliography}

%
%
%

\end{document}